\documentstyle[prb,aps,epsf,psfig,multicol]{revtex}

\newcommand{\ra}{\rangle}

\begin{document}
\draft

\title{Finite-Temperature Transport in Finite-Size Hubbard Rings
in the Strong-Coupling Limit}
\author{N. M. R. Peres}

\address{Departamento de F\'{\i}sica, Universidade de \'Evora,
Rua Rom\~ao Ramalho, 59, P-7001 \'Evora Codex, Portugal}

\author{R. G. Dias }
\address{Departamento de F\'{\i}sica, Universidade de Aveiro, 
P-3800 Aveiro, Portugal}

\author{P. D. Sacramento}
\address{Departamento de F\'{\i}sica and CFIF, Instituto Superior 
T\'ecnico, Av. Rovisco Pais, 1049-001 Lisboa, Portugal}

\author{J. M. P. Carmelo}
\address{N.O.R.D.I.T.A., Blegdamsvej 17, DK-2100 Copenhagen \O,
Denmark and\\ 
Departamento de F\'{\i}sica, Universidade de \'Evora,
Rua Rom\~ao Ramalho, 59, P-7001 \'Evora Codex, Portugal}

\date{\today}

\maketitle

\begin{abstract}
We study the current, the curvature of levels, and the finite 
temperature charge stiffness, $D(T,L)$, 
in the strongly correlated limit, $U\gg t$, for Hubbard rings
of $L$ sites,
with $U$ the on-site Coulomb repulsion and $t$ the hopping
integral. Our study
is done for finite-size systems and any band
filling. Up to order $t$ we derive our results following
two independent approaches, namely, using the solution	provided
by the Bethe ansatz and the solution provided by an algebraic
method, where the electronic operators are represented in a slave-fermion
picture. We find that, in the $U=\infty$ case, the finite-temperature
charge stiffness
is finite for electronic densities, $n$, smaller than one. 
These results are essencially
those of spinless fermions in a lattice of size $L$, apart  from small
corrections
coming from a statistical flux, due to the spin degrees of freedom.
Up to order $t$, the Mott-Hubbard gap is $\Delta_{MH}=U-4t$, 
and we find that $D(T)$ is finite for 
$n<1$, but is zero at half-filling. This result comes from
the effective flux felt by the holon excitations, which, due to
the presence of doubly occupied sites, is renormalized to 
 $\Phi^{eff}=\phi(N_h-N_d)/(N_d+N_h)$, and which is zero at half-filling,
with $N_d$ and $N_h$ being the number of doubly occupied and empty lattice
sites, 
respectively. 
Further, for half-filling, the current transported by any 
eigenstate of the system
is zero and, therefore, $D(T)$ is also zero.
\end{abstract}
\vspace{0.3cm}
\begin{multicols}{2}
\narrowtext
%%%%%%%%%%%%%%%%%%%%%%%%%%%%%%%%%%%%%%%%%%%%%%%%%%%%%%%%%%%%%%%%%%%%%%%%%%%%
%%%%%%%%%%%%%%%%%%%%%SECTION%%%%%%%%%%%%%%%%%%%%%%%%%%%%%%%%%%%%%%%%%%%%%%%%
\section{Introduction}

The transport properties of strongly correlated electrons in 
low-dimensional conductors has been a subject of experimental and 
theoretical interest for over twenty years. Low-dimensional conductors 
show large deviations in their transport properties from the usual 
single-particle description. This suggests that electronic correlations 
play an important role in these systems 
\cite{Jacobsen,Donovan,Danilo,Degiorgi,Kim,Mori,Kobayashi},
even if the correlations  are small \cite{Mori}. Solvable 
one-dimensional many-electron models such as the Hubbard chain are 
often used as an approximation for the study of the properties of 
quasi-one-dimensional conductors.
Although the Hubbard chain has been 
diagonalized long ago \cite{Lieb,Takahashi72}, the involved form of 
the Bethe-ansatz (BA) wave function has prevented the calculation of 
dynamic response functions, these including the charge-charge and 
spin-spin response functions and their associated conductivity spectra. 
Information on low-energy expressions for correlation functions
can be obtained by combining BA with conformal-field theory
\cite{Frahm}. On the other hand, several approaches using 
perturbation theory \cite{Maldague}, 
bosonization \cite{Schulz90,Gimarchi}, the pseudoparticle formalism 
\cite{Carm4,Carmelo92,Nuno97}, scaling methods \cite{Stafford}, 
and spin-wave theory 
\cite{Horsch} have been used to investigate the low-energy transport 
properties of the model away from half filling and at the 
metal -- insulator transition \cite{Lieb}. 
Limited information on the transport properties at finite 
energies has been obtained by numerical methods 
\cite{Loh,Fye,NunoZFPB}.

Recently, a series of comparative numerical and analytical studies
have explored the differences in the transport properties between
integrable and non-integrable 1D models 
\cite{Castella95,Castella96,Zotos96,Antonio96,Naef97,Narozhny98,Zotos99,Peres99},
at finite temperature.
Most of these studies have dealt with generalizations to
finite-temperature
of Kohn's zero-temperature concepts and approach \cite{Kohn64}.
Zotos and Prelov\v{s}ek  \cite{Zotos96} have
introduced the concepts of {\it ideal
insulator} and {\it ideal conductor} 
at {\it finite} temperatures.
These concepts refer to the 
temperature dependence of the real part of the optical
conductivity $\sigma_r(\omega,T)$, which is given by
	\begin{equation}
	\sigma_r(\omega,T)=2\pi D(T)\delta(\omega)+
	\sigma_{reg}(\omega,T)\,,
	\end{equation}  
where we have taken $\hbar=e^2=1$, $e$ is the electron charge,
$\omega$ is the frequency of the electric field, and $T$ is the
temperature.
The quantity $D(T)$ is the charge stiffness 
and characterizes the
response of the
system to a static electric field, within linear response theory. 
According to Kohn's
zero-temperature criterion, the value of $D(0)$ can be used to 
distinguish between an
ideal insulator --$D(0)=0$ -- and an ideal conductor -- $D(0)\neq 0$. 
The quantity
$\sigma_{reg}(\omega,T)$ is the regular part of the conductivity and
describes the absorption of light of finite frequency $\omega$ by the
system.

In this work we are concerned with the one-dimensional Hubbard model.
The response of
the energy eigenvalues, 
$E_m$, of the model to an external flux $\phi={\cal A}L$ (${\cal A}$ is the 
vector
potencial along the chain) 
piercing the ring can be used
to determine $D(T)$.

The charge stiffness $D(T,L)$,
for a finite-size system of length $L$, can be evaluated as a
thermodynamic quantity \cite{Castella95,Castella96,Zotos96} and is given
by
\begin{equation}
D(T,L)=\frac 1{ZL}\sum_m D_me^{-E_m/T}
	\,,
\end{equation}
and
\begin{equation}
2D_m=\left.\frac 
{d^2\,E_m}{d(\phi/L)^2}\right\vert_{\phi=0}\,,
\end{equation}
where $Z$ is the partition function and $D_m$ is the curvature of
the eigenenergy $E_m$. This equation is derived using finite temperature
linear response, and was obtained for the first time by
Castella,  Zotos, and Prelov\v{s}ek in connection with the problem
of diffusive or ballistic transport of a particle in a fermionic bath
\cite{Castella95}.

Another equation for the charge stiffness $D(T)$ can be derived starting
with the partition function $Z(\phi)$. Taking the second derivative
of the free energy ${\cal F}=-T\ln Z(\phi)$ in order to $\phi/L$ 
we arrive to an alternative relation
for $D(T)$
\begin{equation}
	D(T,L)=D_1(T,L)+D_2(T,L)\,,
\label{dt}
\end{equation} 
where
\begin{eqnarray}
	D_1(T,L)&=&\frac 1{2L}\left.
	\frac{d^2\,{\cal F}}{d(\phi /L)^2}\right\vert_{\phi=0}
\nonumber\\ 
	D_2(T,L)&=&\frac 1{2TLZ}\sum_m \left ( j_m \right )^2
e^{-E_m/T}
\nonumber\,,
\end{eqnarray}
and
\begin{equation}
 j_m=-\left.\frac{
d\,E_m}{d(\phi/L)}\right\vert_{\phi=0}
\end{equation}	
The terms
$D_1$ and $D_2$ represent 
the thermodynamic and the current contributions,
respectively. The thermodynamic contribution, $D_1(T)$, vanishes in the
thermodynamic limit \cite{Zotos96}.

Zotos and  Prelov\v{s}ek \cite{Zotos96}, based 
on numerical work done in small sistems,
conjectured, for integrable systems -- namely the spinless-fermion and the
Hubbard models --, that $D(T)$ should be zero if $D(0)$ is also zero, but
finite otherwise. Very recently, Zotos \cite{Zotos99} has shown that this
is so
for the isotropic Heisenberg model, and some of the present authors have shown
\cite{Peres99}
that Zotos conjecture also holds for the $t-V$ model, in the strong
interacting limit
$V\gg t$, with $V$ the nearest-neighbour Coulomb integral. 

Very recently,
Kirchner, Evertz, and Hanke have studied $D(T,L)$ 
using Quantum Monte Carlo simulations
\cite{Kirchner98}. They have found that
some conjectures first proposed by Zotos and
Prelov\v{s}ek \cite{Zotos96}, regarding the issue of
ideal-insulator behavior in one-dimensional integrable systems,
should not be true, in qualitative 
agreement with a thermodynamic Bethe-ansatz
\cite{Takahashi97}
calculation performed by Fujimoto and Kawakami \cite{Kawakami98}.

At finite values of $L$ and at zero temperature,
the full calculation of $D(0,L)$, for any value of $U$,
was first considered by Stafford, Millis and
Shastry 
\cite{Stafford}. These authors
derived scaling relations for $D(0,L)$. In particular, at
half-filling and for the
strong coupling regime, $U\rightarrow \infty$, 
we have $D(0,L)\simeq (-1)^{L/2+1}L^{1/2}AU\exp[-L/\xi(U)]$,
where $A$ is a constant.   
The study of persistent currents $j_0(\phi/L)$ in the ground state for 
finite
size Hubbard rings and its relation with mesoscopic transport in 
arrays of quantum dots has been
considered by Kusmartsev, Weisz, Kishore, and Takahashi \cite{Kusmartsev},
and by Yu and Fowler \cite{Yu92}.

It is interesting  to remark that,
as in a Fermi liquid, in the thermodynamic limit,
$D(0,\infty)$ can also be derived by means of kinetic equations
for the elementary excitations \cite{Carmelo92}.
 
In this paper we present an analytical study of $j_m$, $D_m$, and $D(T,L)$
for 
Hubbard rings of different sizes.  
Our results refer to infinite
and very large values of $U$. In addition to the BA based representation,
we also use a slave-fermion representation \cite{Dias92},
both of them leading to the same physical results.  

The paper is organized as follows. In Sec. \ref{baphi} we present the
solution of the BA equations with twisted boundary conditions. We use
these
equations to compute the mean value of the current operator in any
eigenstate of the model and the corresponding 
curvature of the energy level. These
results allow us to determine the finite-temperature behavior of the
charge stiffness. We do our calcutions in the $U=\infty$ and $U\gg t$
cases. In Sec. \ref{algeb} we rederive the results of Sec \ref{baphi},
using an algebraic approach, where the electronic operators are
described in a slave-fermion representation, and the same physical
quantities
are computed up to order $t$. In Sec. \ref{semi} we make a comparison
between two charge gapped systems, at half-filling: 
the Mott-Hubbard insulator (represented by
the Hubbard model) and the band insulator 
(represented by a dimerized lattice model). In Sec. \ref{conclu} we
present
our conclusions.   

%%%%%%%%%%%%%%%%%%%%%%%%%%%%%%%%%%%%%%%%%%%%%%%%%%%%%%%%%%%%%%%%%%%%%%%%%%%%
%%%%%%%%%%%%%%%%%%%%%SECTION%%%%%%%%%%%%%%%%%%%%%%%%%%%%%%%%%%%%%%%%%%%%%%%%
%%%%%%%%%%%%%%%%%%%%%%%%%%%%%%%%%%%%%%%%%%%%%%%%%%%%%%%%%%%%%%%%%%%%%%%%%%%%
\section{Charge stiffness $D(T)$ for a finite-size system}
\label{baphi}

\subsection{Bethe-ansatz equations}

In this section we study the charge stiffness at finite
temperature for 
one-dimensional Hubbard rings of finite-size $L$.  
The Hamiltonian for the Hubbard model is given by

 	\begin{eqnarray}
	\hat{H} &=& -t\sum_{j,\sigma}
	[c_{j\sigma}^{\dag}c_{j+1\sigma} + H. c.]
	+U\sum_{j} \hat{n}_{j,\uparrow}
	\hat{n}_{j,\downarrow}\, ,
	\label{hamilt}
	\end{eqnarray}
where $c_{j\sigma}^{\dag }(c_{j\sigma})$ creates (annihilates)
one electron with spin $\sigma$ (here and when used as operator 
index, $\sigma = \uparrow,\downarrow$, and $\sigma=\pm1$ otherwise),
$\hat{n}_{j,\sigma}=c_{j\sigma}^{\dag}c_{j\sigma}$ is the number
operator at site $j$, 
and $c_{L+1\sigma}=c_{1\sigma}$.

In the presence of a time-dependent vector potencial ${\cal A}(t)$ the 
hopping amplitude $t$ changes according to the well-known Peierls
substitution

\begin{eqnarray}
	t_{i+1,i}&&\rightarrow t_{i+1,i}({\cal A}(t))\nonumber\,, \\
	t_{i+1,i}&&({\cal A}(t))=t_{i+1,i}\exp[-ie(R_{i+1}-R_i){\cal
A}/\hbar c]\,,
\label{phase}
\end{eqnarray}
where $R_i$ is the position of the site $i$ on the lattice.
It has been possible to solve the Hamiltonian (\ref{hamilt}) 
with the additional hopping phase $\exp([-ie(R_{i+1}-R_i){\cal A}/\hbar
c])$.
For convenience we write ${\cal A}=\hbar c\phi/(a e L)$, where the lattice
spacing
$a$ is given by $a=R_{i+1}-R_i$.
More generaly, a spin-dependent ${\cal A}_{\sigma}$ vector potencial 
can be introduced and the model (\ref{hamilt}) can be solved
by means of the coordinate BA either with twisted or toroidal 
boundary conditions, both approaches giving essentially the 
same results \cite{Shastry,Martins}. One obtains the energy 
spectrum of the model parameterized by a set of numbers 
$\{k_j,\Lambda_\delta\}$ which are solution of the BA
interaction equations given by

	\begin{equation}
	e^{ik_j L}=e^{i\phi_{\uparrow}}\prod_{\delta=1}^{N_\downarrow}
	\frac{\sin k_j-\Lambda_\delta+iU/4}{\sin k_j-\Lambda_\delta-
	iU/4}\, ,
	\hspace{0.5cm} (j=1,\ldots,N)\, ,
	\label{inter1}
	\end{equation}
and
\begin{eqnarray}
	e^{-i(\phi_{\downarrow}-\phi_{\uparrow})}
	\prod_{j=1}^{N}
	\frac{\sin k_j-\Lambda_\delta+iU/4}{\sin k_j-\Lambda_\delta-iU/4}
	=\nonumber\\
	-\prod_{\beta=1}^{N_\downarrow}
	\frac{\Lambda_\beta-\Lambda_\delta+iU/2}
	{\Lambda_\beta-\Lambda_\delta-iU/2}\, ,
	\hspace{0.5cm} (\delta=1,\ldots,N_\downarrow)\, .
\label{inter2}
\end{eqnarray}

The above equations have both real and complex solutions for 
the rapidities $k_j$ and $\Lambda_{\beta}$.
However, previous studies of the $\phi_\sigma\neq 0$ problem 
\cite{Shastry,Martins} have only considered the real
BA rapidities solutions of Eqs. (\ref{inter1}) and (\ref{inter2})
which refer to the low energy excitation spectra. 
The general solution of Eqs. (\ref{inter1}) and (\ref{inter2}) for a
finite
system of size $L$ has been
obtained previously in the literature \cite{NunoZFPB,Kawakami98}
and is given by
%%%%%%%%%%%%%%%%%%%%%%%%%%%%%%%
\begin{eqnarray}
	k_jL &=&
	2\pi I_j^c+\phi
	-\sum_{\gamma }\sum_{j'=1}^{N_{s,\gamma}}
	2\tan^{-1}\left(\frac{\sin k_j/u-R_{s,\gamma,j'}}{\gamma+1}
\right)
		\nonumber \\
	&-& \sum_{\gamma >0}\sum_{j'=1}^{N_{c,\gamma}}
	2\tan^{-1}
	\left(\frac{\sin k_j/u-R_{c,\gamma,j'}}{\gamma}\right)\,,
	\label{tak1}
\end{eqnarray}
\begin{eqnarray}
	2L \, \sin^{-1}\left(u\sqrt{\gamma^2+(R_{c,\gamma,j}+1/u)^2}\right.
	\nonumber\\
	\left.-
	u\sqrt{\gamma^2+(R_{c,\gamma,j}-1/u)^2}\right)
	= 2\pi I^{c,\gamma}_j
	- 2 \gamma \phi-\nonumber\\
	-\sum_{j'=1}^{N_c}2\tan^{-1}\left(\frac{\sin
k_{j'}/u-R_{c,\gamma,j}}
	{\gamma}\right)\nonumber\\
	+\sum_{\gamma'>0}\sum_{j'=1}^{N_{c,\gamma'}}
	\Theta_{\gamma,\gamma'}(R_{c,\gamma,j}-R_{c,\gamma',j'})\, ,
\label{tak2}
\end{eqnarray}
and
\begin{eqnarray}
	&&\sum_{j'=1}^{N_c}2\tan^{-1}\left(\frac{R_{s,\gamma,j}-\sin
k_{j'}/u}
	{1+\gamma}\right)=
	\nonumber \\
	&& 2\pi I^{s,\gamma}_j+
	\sum_{\gamma'}\sum_{j'=1}^{N_{s,\gamma'}}
	\Theta_{\gamma+1,\gamma'+1}(R_{s,\gamma,j}-R_{s,\gamma',j'})\, .
	\label{tak3}
\end{eqnarray}
In the above equations we have considered the case 
$\phi_{\sigma}=\phi_{-\sigma}$ and $u=U/(4t)$.

The functions $\Theta_{\gamma,\gamma'}(x)$ [and 
$\Theta_{\gamma+1,\gamma'+1}(x)$] of Eqs.
\,(\ref{tak1}), (\ref{tak2}), and (\ref{tak3}) are
defined in Ref. \onlinecite{Nuno97}, and are sums of
$\tan^{-1}(x)$. The following definitions
for the real part of the rapidities, 
$\Lambda_\alpha^{n+1}/u=R_{s,\gamma,j}$ (with $n+1=\gamma$ and 
$\alpha=j$), $\Lambda_\alpha^{' \, n}/u=R_{c,\gamma,j}$ 
(with $n=\gamma$ and $\alpha=j$), and $\gamma=1,2,\ldots,\infty$ 
for the $N_{c,\gamma}$ sums and $\gamma=0,1,2,\ldots,\infty$ for 
the $N_{s,\gamma}$ sums, allow us to recover Takahashi's 
formulae for $\phi=0$ \cite{Takahashi72}. Here and often below
we use the notation $c\equiv c,0$, which allows the $c,\gamma$
sums to run over $1,2,3,\ldots$ . 

The relevant numbers $I_j^c$, $I^{c,\gamma}_j$, and 
$I^{s,\gamma}_j$ which appear in going from Eqs.\,(\ref{inter1}) 
and (\ref{inter2}) to Eqs.\,(\ref{tak1}), (\ref{tak2}), and 
(\ref{tak3}) are the quantum numbers which describe
the Hamiltonian eigenstates. In Table \ref{class} we
give a classification of the type of excitations
described by these quantum numbers.

The numbers $I_j^{\alpha,\gamma}$ can be integers or half-odd integers
\cite{Takahashi72} if the numbers ${\bar N}_{\alpha,\gamma}$ are even or
odd,
respectively, where

\begin{eqnarray}
{\bar N}_{c}=\sum_{\gamma=0}N_{s,\gamma}+\sum_{\gamma=1}
N_{c,\gamma}\nonumber\,,\\
{\bar N}_{c,\gamma}=1+L-N+N_{c,\gamma}\nonumber\,,\\
\hspace{.5cm}{\bar N}_{s,\gamma}=1+N-N_{s,\gamma}\,.
\end{eqnarray} 
The spacing between adjacent quantum numbers  $I^{\alpha,\gamma}_j$, 
is always one, and independent of the value of the interaction $U$.
It is therefore natural to interpret $q_j^{\alpha,\gamma}=2\pi
I_j^{\alpha,\gamma}/L$
as a momentum  \cite{Nuno97}, 
and the rapidities $R_{\alpha,\gamma,j}$ and $k_j$ as
functions
of $q_j^{\alpha,\gamma}$.
The total number of $\sigma$ electrons, $N_{\sigma}$,
is given by the constraints

\begin{equation}
N_\downarrow = A_c 
+ A_s\, ,
\label{nbaixo}
\end{equation}
and

\begin{equation}
N = N_{c} + 2A_c\,,
\label{nc}
\end{equation}
where
\begin{eqnarray}
 A_c&=&\sum_{\gamma=1}\gamma N_{c,\gamma}\nonumber\,\\
 A_s&=&\sum_{\gamma=0}(\gamma+1)N_{s,\gamma}\,,
 \end{eqnarray}
and the numbers $I^{c}_j$, $I^{c,\gamma}_j$, and
$I^{s,\gamma}_j$ belong to the intervals

\begin{eqnarray}
 	\vert I^{c}_j \vert &&< \frac L{2}\nonumber\,,\\
	\vert I^{c,\gamma}_j \vert &&< \frac 1{2}
	(L-N+2A_c-T_c^{\gamma})\nonumber\,,\\
	\vert I^{s,\gamma}_j \vert &&< \frac 1{2}
	(N-2A_c-T_s^{\gamma})\,,
\end{eqnarray}
where  $T_{\alpha}^{\gamma}$ (with $\alpha=c,s$) are given by

\begin{eqnarray}
 	 T_c^{\gamma}&=&\sum_{\gamma '=1}
  	t^c_{\gamma,\gamma '}N_{c,\gamma '}\nonumber\,\\
 	T_s^{\gamma}&=&\sum_{\gamma '=0}t^s_{\gamma,\gamma '}N_{s,\gamma
'}\,,
\end{eqnarray}
with $t^c_{\gamma, \gamma '}=2 \min(\gamma, \gamma ')-\delta_{\gamma,
\gamma '}$
and $t^s_{\gamma, \gamma '}=2 \min(\gamma+1, \gamma '+1)-\delta_{\gamma,
\gamma '}$.

All the eigenstates considered in the
BA solution
of the model are described
by the different occupancies of the quantum numbers $I_j^{\alpha,\gamma}$. 
For example,
the ground state is described \cite{Nuno97} by a compact symmetric
occupancy around 
the origin of the numbers $I_j^c$ and $I_j^{s,0}$, and by zero 
occupancy for the numbers $I^{c,\gamma}_j$ and 
$I^{s,\gamma>0}_j$.
The general situation is given in Table \ref{stateg}. 

The energy and momentum eigenvalues are given by \cite{Takahashi72,Nuno97}

\begin{eqnarray}
	E(L,\phi,U,N_{\sigma})=-2t\sum_{j=1}^{N_c}\cos k_j+\nonumber\\
	4t\sum_{\gamma=1}\sum_{j=1}^{N_{c,\gamma}} Re
	\sqrt{1-u^2[R_{c,\gamma,j}-i\gamma]^2}\,,
\label{energy}
\end{eqnarray}
and
\begin{eqnarray}
	P&=&\sum_{j=1}^{N_c}\frac {2\pi}{L}I_j^c+
	\sum_{\gamma}\sum_{j=1}^{N_{s,\gamma}}\frac {2\pi}{L}I_j^{s,\gamma}
	-\sum_{\gamma}\sum_{j=1}^{N_{c,\gamma}}\frac {2\pi}{L}I_j^{c,\gamma}+
	\nonumber\\
	&+&\frac N L \phi +\pi\sum_{\gamma}N_{c,\gamma}\,.
\label{momentum}
\end{eqnarray}
It is interesting to remark that the flux contribution $P_{\phi}$ for the total
momentum $P$ differs from its  contribution
in the independent electronic problem, 
for which we would obtain a contribution
of the form $P_{\phi}=\phi N/(2L)$. 
The general solution of Eqs. (\ref{tak1})-(\ref{tak3}) for arbitrary
system size $L$, electron numbers $N_{\sigma}$, and Coulomb interaction 
$U$ is a very difficult problem. Close to and at half-filling 
Stafford {\it et al.} computed the 
zero temperature
charge stiffness 
for a system of size $L$ \cite{Stafford}. To compute the charge stiffness
at finite temperature we need to compute the mean value
of the current operator or the curvature of levels
of all the eigenstates of the model and
not only that of the ground state. 
Below, we solve the problem explicitly for a system
of size $L$ in the limits $U=\infty$  and $U \gg t$.
For simplicity we shall consider the case where the quantum numbers
$I^{\alpha,\gamma}_j$ are symmetrically distributed arround the origin.
%%%%%%%%%%%%%%%%%%%%%%%%%%%%%%%%%%%%%%%%%%%%%%%%%%%%%%%%%%%%%
\subsection{The $U=\infty$ case}
\label{infinity}

At $U=\infty$ the charge strings have infinite energy and, therefore, drop
out
of the problem. In physical terms this means 
that the states with one or more doubly occupied sites are not permited.
The rapidites $R_{s,\gamma,j}$ decouple from the charge degrees of freedom
and
all the spin excitations are degenerate
\cite{Dias92,Woynarovich82,Ogata90}. 
The  $s,0$ spinon and the $s,\gamma >0$ spin-string excitations 
have a flat dispersion relation. 
The equations (\ref{tak1})-(\ref{tak3}) reduce to

\begin{eqnarray}
	k_j^{\infty}L =
	2\pi I_j^c+\phi
	&+&\sum_{\gamma }\sum_{j'=1}^{N_{s,\gamma}}
	2\tan^{-1}\left(R_{s,\gamma,j'}^{\infty}/(\gamma+1) \right)
	\nonumber \\
	&+&\sum_{\gamma }\sum_{j'=1}^{N_{c,\gamma}}
	2\tan^{-1}\left(R_{c,\gamma,j'}^{\infty}/\gamma \right)\,,
	\label{tak1inf}
\end{eqnarray}

\begin{eqnarray}
	(L-N_c)2\tan^{-1}\left(R_{c,\gamma,j}^{\infty}/
	\gamma\right) =
	\nonumber \\
	= 2\pi I^{c,\gamma}_j-2\gamma \phi+
	\sum_{\gamma'}\sum_{j'=1}^{N_{c,\gamma'}}
	\Theta_{\gamma,\gamma'}
	(R_{c,\gamma,j}^{\infty}-R_{c,\gamma',j'}^{\infty})\, ,
	\label{tak2inf}
\end{eqnarray}
and
\begin{eqnarray}
	N_c2\tan^{-1}\left(R_{s,\gamma,j}^{\infty}/
	(1+\gamma)\right) =
	\nonumber \\
	= 2\pi I^{s,\gamma}_j+
	\sum_{\gamma'}\sum_{j'=1}^{N_{s,\gamma'}}
	\Theta_{\gamma+1,\gamma'+1}
	(R_{s,\gamma,j}^{\infty}-R_{s,\gamma',j'}^{\infty})\,.
	\label{tak3inf}
\end{eqnarray}
We see from the structure of Eqs. (\ref{tak1inf})-(\ref{tak3inf})
that the Hubbard model Hilbert space decouples into a product of three
Hilbert subspaces, each one associated with a different Hamiltonian.
The latter are a chain of length $L$ and $N_c$ spinless fermions, an
Heisenberg spin one-half chain of length $N_c$, and  an
Heisenberg spin one-half chain of length $L-N_c$. This is, however, a very
delicated decoupling in what regards the thermodynamic properties. Since
in a thermodynamic calculation
$N_c$ must vary between $N_c=L$ and $N_c=0$, the length of the two
Heisenberg chains also varies.

The energy eigenvalues are given by

\begin{equation}
	E(L,\phi,N_{\sigma})=E_m^{\infty}=
	-2t\sum_{j=1}^{N_c=N}\cos k_j^{\infty}\,,
\label{energyinf}
\end{equation}
where the subscript $m$ labels a given eigenstate of the model.
Similar equations to Eqs. (\ref{tak1inf})-(\ref{energyinf}) 
have been derived for the ground state,
 in the study of persistent currents
in finite-size rings \cite{Kusmartsev}. If we use Eqs. (\ref{tak2inf})
and (\ref{tak3inf})
in Eq. (\ref{tak1inf}) we obtain a solution for $k^{\infty}_j$
in terms of the quantum numbers $I^{c}_j$ and $I^{\alpha,\gamma}_j$
only. This solution reads 

\begin{eqnarray}
	k_j^{\infty} =
	\frac {2\pi}{L} I_j^c&+&\frac{\phi}{L}
	\frac{L-N_c-\sum_{\gamma}2\gamma N_{c,\gamma}}{L-N_c}
	 + \frac {2 \pi}{L N_c}
	\sum_{\gamma }\sum_{j=1}^{N_{s,\gamma}}I^{s,\gamma}_j 
	\nonumber\\
	&+&\frac {2 \pi}{L (L-N_c)}
	\sum_{\gamma }\sum_{j=1}^{N_{c,\gamma}}I^{c,\gamma}_j\,.
	\label{tak1inftot}
\end{eqnarray}
Equation (\ref{tak1inftot}) shows that the spin degrees of freedom are
still coupled
to the charge excitations through the quantum numbers $I^{s,\gamma}_j$.
These
act as a ficticious flux piercing the ring of spinless fermions. The fact
the charge and spin degrees of  freedom are not completely decoupled
introduces
interesting statistical consequences in the classification of
the excitations according to the Haldane criterium 
\cite {Nuno97,PRL98,Haldane91}.

The eigenstates with non-zero occupancy of the
numbers $N_{c,\gamma}$ have infinite energy (for $U=\infty$)
and therefore 
they drop out of the problem. That is, $k_j^{\infty}$
has the simpler form

\begin{equation}
	k_j^{\infty} =
	\frac {2\pi}{L} I_j^c+\frac{\phi}{L}
	 + \frac {2 \pi}{L N_c}
	\sum_{\gamma }\sum_{j=1}^{N_{s,\gamma}}I^{s,\gamma}_j\,.
	\label{tak1infb}
\end{equation}

The curvature of levels $D_m(L)$ for an eigenstate $m$ is given by

\begin{equation}
	D_m(L)=\frac{1}{2L}\frac {d\,^2 E_m}{d\,(\phi/L)^2}=
	\frac 1{L}\sum_{\{I^{c}_j\}_m}d(I^{c}_j)\,,
\label{dofm}
\end{equation}
where the sum is over the configuration of quantum numbers 
$I^{c}_j$ defining the eigenstate $m$ and $d(I^{c}_j)$ is given by

\begin{eqnarray}
	d(I^{c}_j)&=&t\cos k_j \left ( 
	\frac {d\,k_j}{d\,(\phi /L)} \right )^2 +
	t\sin k_j \frac {d^2\,k_j}{d\,(\phi/L)^2}\nonumber\\
&=&t\cos k_j^{\infty}\,.
\label{dtinfty}
\end{eqnarray}
From Eq. (\ref{tak1infb}) we expect
the behavior of the charge stiffness
$D(T,L)$ to be essencialy the same as for free spinless fermions. Small
differences are to be expected due to the statistical flux term
$\phi^{stat}_s$

\begin{equation}
	\phi^{stat}_s=\frac{2 \pi}{L N_c}
	\sum_{\gamma }\sum_{j=1}^{N_{s,\gamma}}I^{s,\gamma}_j\,. 
\label{fstat}
\end{equation}
If we consider the special case of a fully polarized system
the statistical flux term vanishes and the system
is equivalent to
free spinless fermions in a ring of size $L$. 

\begin{figure}[f]
\epsfxsize=6.5 cm 
\epsfysize=8.0 cm
\centerline{\epsffile{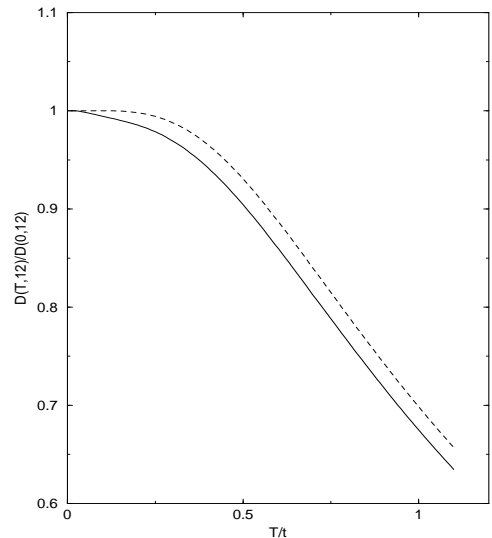}}
\caption{The charge stiffness 
$D(T,12)/D(0,12)$ for 
$L=12$,  $N_{\uparrow}=N_{\downarrow}=3$. The result for free
spinless fermions is the dotted line, the full line is the
result for the $U=\infty$ Hubbard model.}
\label{figinfty}
\end{figure}

To illustrate
the differences between free spinless fermions
and the Hubbard model at $U=\infty$ when $\phi^{stat}_s\neq 0$
we plot in Figure \ref{figinfty} $D(T,L)$ for the $U=\infty$
Hubbard and for the spinless free-fermion models in a ring of $L=12$ for 
$N_{\uparrow}=N_{\downarrow}=3$. For this case we 
have, in the Hubbard model, two types
of spin-string excitations, those with $\gamma=1$ and $\gamma=2$. The
details 
of the states available are given in Table \ref{uinftydet}. The results of
$D(T,L)$ in Figure \ref{figinfty}
are normalized by the zero-temperature charge stiffness $D(0,L)$ 
which, for $N$ even, is given for the Hubbard model by 

\begin{equation}
D_U(0,L)=t\frac{\sin (\pi N/L)}{\sin (\pi /L)}
\frac {1}{L}\,,\hspace{.5cm}
(U=\infty)\,,
\end{equation}
and for free spinless fermions by
\begin{equation}
D_0(0,L)=D_U(0,L)\cos{\pi \over L}\,,\hspace{.3cm}
(U=0)\,.
\label{zerodt}
\end{equation}
From Eqs. (\ref{dofm}) and (\ref{dtinfty}) it is clear that at
half-filling the charge stiffness $D(T)$ is zero
at any temperature, since the only available
state is the ground state (apart from a
large massive spin degeneracy). In this limit the spin degrees 
of freedom $s,\gamma$  introduce a large degeneracy but 
they do not contribute to the transport of charge.

We see from Eq. (\ref{dt}) that the charge stiffness
can be obtained from a thermodynamic contribution
plus a contribution from the thermal 
average of the square of the mean value
of the current operator, $j_m$,  
that we have denoted by $D_2(T,L)$.

\begin{figure}[f]
\epsfxsize=6.5 cm 
\epsfysize=8.0 cm
\centerline{\epsffile{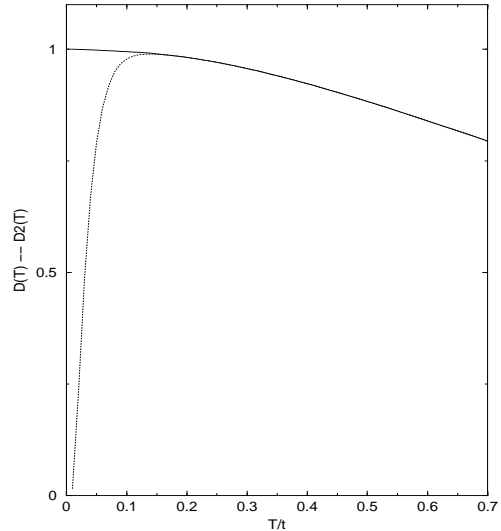}}
\caption{The charge stiffness, 
$D(T,L)/D(0,L)$ (line), and the thermal average
of $(j_m)^2$, $D_2(T,L)/D(0,L)$ 
(dots), for the Hubbard model. The system size is
$L=18$, the number
of electrons is $N_{\uparrow}=N_{\downarrow}=3$,
and the on-site Coulomb repulsion is $U=\infty$.}
\label{figdtd2}
\end{figure}

In the limit $U=\infty$ the mean value of the current operator is easily
computed in a finite-size system and reads

\begin{equation}
	j_m^{\infty}(L)=-\frac 1{L}\sum_{\{I^{c}_j\}_m}2t\sin
k_j^{\infty}\,.
\label{jmeaninfty}
\end{equation}
In Figure \ref{figdtd2} 
we present the results for
$D(T,L)$ and $D_2(T,L)$, for $L=18$ and
$N_{\uparrow}=N_{\downarrow}=3$. We see that
for $T> t/10$ $D(T,L)$ and $D_2(T,L)$
coincide ($T^{\ast}\simeq t/10$ being a crossover temperature
for $L=18$). 
On the other hand, at low $T$ 
$D_2(T,L)$ is much smaller 
than $D(T,L)$, 
for finite-size
sytems.
We have checked that
$D(T,L)$ and $D_2(T,L)$ 
coincide at lower and lower temperatures, as $L$ increases.
In the thermodynamic limit  the
contribution  $D_1(T)$ from the free energy vanishes and $D(T)=D_2(T)$.
Similar results have been obtained by
Narozhny, Millis, and Andrei  for the spinless fermion
model \cite{Narozhny98}, 
by exact diagonalization of small
clusters (up to $L=14$).

The behavior discussed above can be easily understood. 
For a given $L$, as $T$
increases the contribution of the groundstate to the thermodynamic
average decreases. Since $j_0$ (the current in the groundstate) is zero, 
$D_2(T)$
eventually has to start decreasing to zero 
at some crossover temperature, where the thermodynamic weight
of the groundstate  dominates  the Boltzmann average. 
When $L$ increases the number of low energy
states above the GS also increases. This reduces the thermodynamic 
importance of the $j_0$ to the thermodynamic
average and the crossover temperature $T^{\ast}$ is shifted to
lower values.

\begin{figure}[f]
\epsfxsize=6.5 cm 
\epsfysize=8.0 cm
\centerline{\epsffile{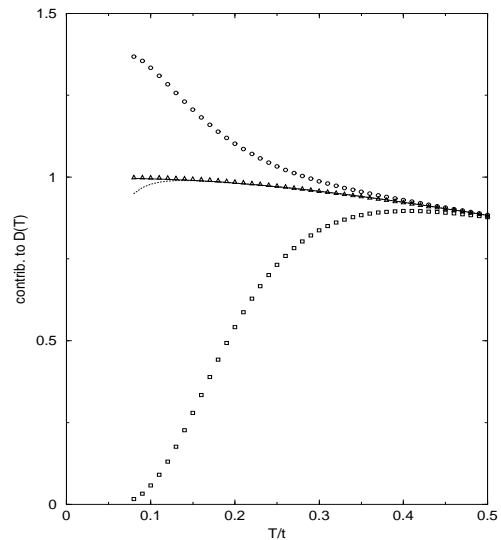}}
\caption{Contributions of the several types of spin excitations 
to $D(T)$ and $D_2(T)$. 
The notation is the following: $D(T,18)$ (full line); $D_2(T,18)$ (dotted
line);
$D(T,18)$ including the $s,0$ excitations only (triangles);
$D_2(T,18)$ including the $s,0$ excitations  only (squares);
$D_2(T,18)$ including the $s,0$ and $s,1$ excitations only (circles).
The system size is
$L=18$, the number
of electrons is $N_{\uparrow}=N_{\downarrow}=3$ and the 
on-site Coulomb repulsion is $U=\infty$.}
\label{figd2spin}
\end{figure}

We now show that the results for  $D_2(T,L)$, away from 
half-filling, for finite-size systems, and low temperature,
are strongly dependent on the
fact that all the spin excitations have been taken
into account. This contrasts with the behavior
of $D(T,L)$ based on the curvature of levels
given by Eq. (\ref{dtinfty}). In Figure 
\ref{figd2spin} we plot the contributions to $D(T)$
due to the several spin excitations -- spinons and
spin strings -- using the curvature of
levels and the mean value of the square of the 
current operator.
From Figure \ref{figd2spin} we see that if we consider
only the spinon $s,0$ excitations the contribution
from these states to $D(T,L)$ (triangles) almost coincides with
the full $D(T,L)$ in the whole range of temperatures. 
On the other hand,
the contribution to $D_2(T,L)$ from the same set of states is very small
at low temperatures. If the
$s,1$ spin-string excitations are included, we see
that $D_2(T,L)$ is now above $D(T,L)$ for low $T$.
As the system size increases the contribution
from the spin excitations decreases.

It is interesting to remark that the electric charge
$e_{c}$ transported by one $c$ excitation is given by $e_{c}=-1$, 
at $U=\infty$. 
These $c$ excitations are the $c$ pseudoparticles of Refs.
\onlinecite{Carm4,Carmelo92}, the holons being their holes
\cite{Nuno97}. Therefore, they transport the same electric 
charge as the original electrons, whereas their holes, the
holons, transport minus the charge of an electron.

%%%%%%%%%%%%%%%%%%%%%%%%%%%%%%%%%%%%%%%%%%%%%%%%%%%%%%%%%%%%%%%%%%%%%%%%%%%%%%%%%%
\subsection{The $U\gg t$ case: ${\cal O}(t)$ and ${\cal O}(t^2/U)$
corrections}
\label{uggt}
In the subsection \ref{infinity} we 
have considered the case of $U=\infty$ where
charge-string excitations are not allowed. In this section we consider the
case
where $U\gg t$ and we solve Eqs. (\ref{tak1}), (\ref{tak2}), (\ref{tak3}),
and
(\ref{energy}) to order $t^2/U$. In this case, $c,\gamma$-excitations can
take
place, since their energies are no longer
infinity. In Table \ref{tabUt1} we list the type of excitations
that can take place in the system for $L=18$ and $N=6$. 

In this limit, the Hilbert space of the Hubbard model can be decomposed
into
several subspaces each of energy of order $\gamma U$ ($\gamma=1,2, \ldots$) 
relatively to the ground
state. In the original lattice this means that we allow for one, two,
...,
doubly occupied sites. These doubly occupied sites are related to the
charge-string excitations $c,\gamma$. For example, in the subspace of
energy
$U$ we can have only one $c,1$ string excitation and in the subspace of
energy
$2U$ we can have two $c,1$ or one $c,2$ string excitations, and so on. On
the other
hand, the conservation of the electron numbers in these excitations,
imposed by 
Eqs. (\ref{nbaixo}) and (\ref{nc}), show that the number
of occupied $I^{c}_j$ is reduced by two and by four, respectively. From
Eq. (\ref{energy})
is simple to show that the energy of these string excitations is given, to
order $t$, by

\begin{equation}
	E_m=-2t\sum_{j=1}^{N_c}\cos k_j^{\infty}
	+U\sum_{\gamma}\gamma N_{c,\gamma}+ {\cal O}(t^2/U)\,.
\label{energyUt}
\end{equation}

Equation (\ref{energyUt}) has first been derived by Woynarovich
\cite{Woynarovich82}, starting from
the ground-state equations of Lieb and Wu \cite{Lieb}. In the limit
$U \gg t$ and to order $t^2/U$ eqs. (\ref{tak1})-(\ref{tak3}) reduce to

\begin{eqnarray}
	k_jL =
	2\pi I_j^c+\phi
	+\sum_{\gamma }\sum_{j'=1}^{N_{c,\gamma}}
	2\tan^{-1}\left(R_{c,\gamma,j'}/\gamma \right)
	\nonumber \\
	-\frac {8t}{U}\sin k_j\sum_{\gamma}\sum_{j'=1}^{N_{c,\gamma}}
	\frac{\gamma}{\gamma^2+(R_{c,\gamma,j'})^2}\nonumber\\
	+\sum_{\gamma }\sum_{j'=1}^{N_{s,\gamma}}
	2\tan^{-1}\left(R_{s,\gamma,j'}/(\gamma+1) \right)
	\nonumber \\
	-\frac {8t}{U}\sin k_j\sum_{\gamma}\sum_{j=1}^{N_{s,\gamma}}
	\frac{\gamma+1}{(1+\gamma)^2+(R_{s,\gamma,j})^2}\,,
\label{tak1Ut}
\end{eqnarray}

\begin{eqnarray}
	(L-N_c)2\tan^{-1}\left(R_{c,\gamma,j}/
	\gamma\right) =
	\nonumber \\
	= 2\pi I^{c,\gamma}_j-2\gamma\phi+
	\sum_{\gamma'}\sum_{j'=1}^{N_{c,\gamma'}}
\Theta_{\gamma,\gamma'}(R_{c,\gamma,j}-R_{c,\gamma',j'})\nonumber\\
	-\frac {8t}{U}\frac{\gamma}{\gamma^2+(R_{c,\gamma,j})^2}
	\sum_{j=1}^{N_{c}}\sin k_j\,,
	\label{tak2Ut}
\end{eqnarray}

and
\begin{eqnarray}
N_c2\tan^{-1}\left(R_{s,\gamma,j}/
(1+\gamma)\right) =
\nonumber \\
= 2\pi I^{s,\gamma}_j+
\sum_{\gamma'}\sum_{j'=1}^{N_{s,\gamma'}}
\Theta_{\gamma+1,\gamma'+1}(R_{s,\gamma,j}-R_{s,\gamma',j'})\nonumber\\
+\frac {8t}{U}\frac{\gamma+1}{(1+\gamma)^2+(R_{s,\gamma,j})^2}
\sum_{j=1}^{N_{c}}\sin k_j\, .
\label{tak3Ut}
\end{eqnarray}
Equations (\ref{tak1Ut}), (\ref{tak2Ut}), and (\ref{tak3Ut}) 
can be solved explicitly
by an iteration procedure introduced first for the
ground-state equations \cite{Kusmartsev}. 
The solution, to order $t^2/U$, it is simple to obtain and
reads

\begin{equation}
	k_j=k_j^{\infty}+\delta_j^s+\delta_j^c\,,
\label{kjasym}
\end{equation}
with $\delta_j^s$ and $\delta_j^c$ given by 
\begin{eqnarray}
	\delta_j^s=&-&\frac{8tB_m^s}{LU}\sin k_j^{\infty}
	+\frac{8tB_m^s}{LUN_c}\sum_{j=1}^{N_c}\sin
k_j^{\infty}\,,\nonumber\\
	\delta_j^c=&-&\frac{8tB_m^c}{LU}\sin k_j^{\infty}
	-\frac{8tB_m^c}{LU(L-N_c)}\sum_{j=1}^{N_c}\sin k_j^{\infty}\,,
\label{delb}
\end{eqnarray} 
and
\begin{equation}
	B_m^s=\sum_{\gamma}\sum_{j=1}^{N_{s,\gamma}}
\frac {1+\gamma}{(1+\gamma)^2+(R_{s,\gamma,j}^{\infty})^2}\,,
\label{bms}
\end{equation}

\begin{equation}
	B_m^c=\sum_{\gamma}\sum_{j=1}^{N_{c,\gamma}}
\frac {\gamma}{\gamma^2+(R_{c,\gamma,j}^{\infty})^2}\,,
\label{bmc}
\end{equation}
and $k_j^{\infty}$ given by Eq. ({\ref{tak1inftot}}). 
The sums in Eqs. (\ref{delb}), (\ref{bms}), and (\ref{bmc})
depend on the distributions of the numbers $I^{\alpha,\gamma}_j$,
that is, on the considered eigenstate.
The $B_m^s$ term
represents
the energy of an isotropic Heisenberg chain, of length $N_c$, with
the spin-string excitations included.  
The $B_m^c$ term has the same form as $B^s_m$ and can
be thought as an {\it effective Heisenberg chain}, of length $(L-N_c)$. At
odds with
$B^s_m$, the excitations represented 
by $B_m^c$ have a charge gap relatively to the ground
state of the system. The smallest gap, $\Delta(L,U)$, is associated with
the 
creation of a single $c,1$ excitation above the ground state 
and, for a finite-size system, is given by
\begin{eqnarray}
	\Delta(L,U)&=&U+4t\cos(N_c \pi/L-\pi/L)\nonumber\\
	&-&\frac{8t^2(N_c-2)}{LU}(B^s_1+1)
	\nonumber\\
	&+& \frac{8t^2}{LU}(B^s_1+1) \frac{\sin(2\pi N_c/L-4\pi/L)}
	{\sin (2\pi/L)}\nonumber\\
	&+&\frac{8t^2}{UL}B^s_0\left(N_c-\frac{\sin(2\pi N_c/L)}
	{\sin (2\pi/L)}\right)\nonumber\\
	&+&\frac{8t^2}{U}+{\cal O}(1/U^2)\,,
	\label{gaplu}
\end{eqnarray}
where $B^s_0$ and $B^s_1$ represent $B_m^s$ for the ground state and for
the
excited state considered, respectively. 
When $L\rightarrow \infty$ and $N_c\rightarrow L$,
we obtain the Mott-Hubbard gap $\Delta_{MH}=U-4t+8t^2\ln 2/U+{\cal
O}(1/U^2)$.

The numbers $R_{c,\gamma,j}^{\infty}$ and $R_{s,\gamma,j}^{\infty}$ are
given 
by the solution of Eqs. (\ref{tak2inf}) and (\ref{tak3inf}), respectively.
The term
in $k_j^{\infty}$ that contains the flux $\phi$ has the form
\begin{equation}
\frac{\phi}{L}
\frac{L-N_c-\sum_{\gamma}2\gamma N_{c,\gamma}}{L-N_c}
=\frac{\Phi^{eff}}{L}.
\label{fluxterm}
\end{equation}
Due to the conservation law (\ref{nc}) it is clear that
at half filling we have $\Phi^{eff}/L=0$ and 
$k_j$ in Eq. (\ref{kjasym}) only depends on $\phi$ through
the $B^c_m$ term. 
In the asymptotic limit $U\gg t$ the sum $\sum_{\gamma}\gamma
N_{c,\gamma}$ is 
identified with the total number of doubly occupied sites $N_d$,
that is
\begin{equation}
	\langle 
	\sum_i {\hat n}_{i,\uparrow} {\hat n}_{i,\downarrow}
	\rangle=
	\frac{\partial E}{\partial U}=\sum_{\gamma}\gamma
N_{c,\gamma}+{\cal O}(1/U^2)\,,
\end{equation}	 
If we defined the number of empty lattice sites as $N_h=L-N+N_d$ we see
that 
$\Phi^{eff}/L$ can be written as
\begin{equation}
	\frac{\Phi^{eff}}{L}=\frac{\phi}
	{L}\frac{N_h-N_d}{N_h+N_d}=\frac{\phi}{L}z.
\label{fluxterm2}
\end{equation}
In addition to the interesting behavior at half-filling, 
we also see from the above equation that if the system is not at 
half-filling 
but doubly occupied sites are allowed then
the flux felt by the $c,0$ excitations is not
the usual $\phi/L$ but the effective flux $\Phi^{eff}/L$.

The corrections of ${\cal O}(t^2/U)$  to the 
energy  (\ref{energyUt}), $E_m^{t^2/U}$, are given
by
\begin{equation}
	 E_m^{t^2/U}=2t\sum_{j=1}^{N_c}\sin
k^{\infty}_j(\delta_j^c+\delta_j^s )+
 	\frac{8t^2}{U}B_m^c\,.
 \label{eoveru}
\end{equation}
In the limit $U\gg t$ it is simple to compute  
the ground-state energy
as a function of $\phi/L$
for a system of any size $L$, which reads
\begin{eqnarray}
E=&-&2t\cos (\phi/L)\frac{\sin(\pi N/L)}{\sin(\pi /L)} 
-\frac{8t^2NB_0^s}{UL}\nonumber\\
&+&\frac{16t^2B_0^s}{ULN}\sin^2(\phi/L)\frac{\sin^2(\pi N/L)}
{\sin^2(\pi /L)}\nonumber\\
&+& \frac{8t^2B_0^s}{UL}\frac{\sin(2\pi N/L)}{\sin(2\pi
/L)}\cos(2\phi/L)\,,
\label{energyu}
\end{eqnarray}
with $B_0^s=N\ln(2)/2$, in the thermodynamic limit.
From equation (\ref{energyu}) it is simple to obtain
$D(0,L)$, for any band filling. 

It is instructive to consider Eq. (\ref{energyu}) in two different cases: (a)  
one electron off half-filling;  (b) at half-filling. The first case gives
for the charge stiffness $D(0,L)=t\delta$, where $\delta=(L-N)/L=1/L$
is the dopping; for the second case we obtain $D(0,L)=0$.
Both these results agree with those of Stafford, Millis and
Shastry \cite{Stafford}. In the second case, we see that $D(0,L)$ has
not corrections of the order $t^2/U$. In fact, the asympotic result shows
that $D(L,U)\propto (4.38/U)^{L-1}$. 

We stress that
the energy (\ref{energyu}) only corresponds to the
true ground state of the model for small values of $\phi$.
When $\phi$ increases level crossing takes place and
therefore the energy (\ref{energyu}) does not
correspond to the absolute ground state of the system.
Furthermore, for $\phi=0$ and $L\rightarrow \infty$, we recover
the results of Carmelo and Baeriswyl 
\cite{Carmelo88} for the
ground-state energy. 

If the corrections (\ref{eoveru}) to the energy are neglected,
the energy eigenvalues depend on $k^{\infty}_j$ only. We then see, from
Eq. (\ref{fluxterm2}),
that the current $d\,E_m/d\,(\phi/L)$ and the curvature 
of levels $d\,^2\,E_m/d\,(\phi/L)^2$ are all proportional to $z$,
that is
\begin{equation}
	\frac{d\,E_m}{d\,(\phi/L)}=-\frac{N_h-N_d}{N_h+N_d}Lj^{\infty}_m\,,
	\label{currz}
\end{equation}
and
\begin{equation}
   	\frac{d\,^2E_m}{d\,(\phi/L)^2}=-z^2E^{\infty}_m\,.
\end{equation}
Away from half-filling, $z$ is finite and $t^2/U$ corrections will
produce no qualitative changes in the results we have
obtained in  subsection \ref{infinity}. We have checked that 
the effect of $t^2/U$ corrections is to increase the value
of $D(T)$ over that of the $U=\infty$ case. This behavior is in agreement
with the  physical picture that  transport of charge is easier when
 electronic correlations are reduced.
\begin{figure}[f]
\epsfxsize=6.5 cm 
\epsfysize=8.0 cm
\centerline{\epsffile{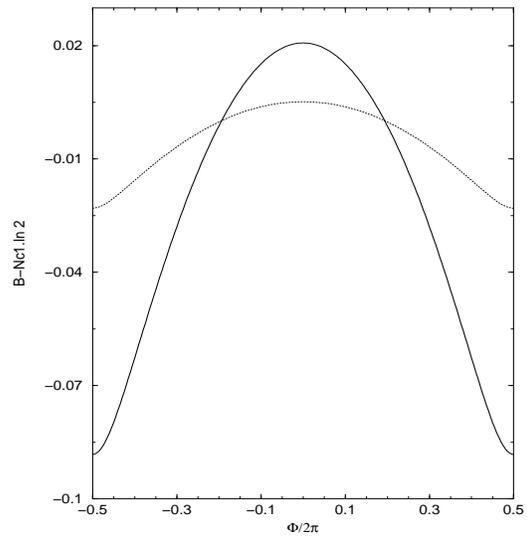}}
\caption{The $B-N_{c,1}\ln 2$ term  as function
of the flux $\phi/(2\pi)$ ($\phi$ is in units of the flux quantum $2\pi$), 
for $N_{c,1}=10$ (solid line) and $N_{c,1}=40$ (dashed line).
It is clear that  $d\,B/d\,\phi$ is zero
at $\phi=0$, and, therefore, the corresponding current matrix element
$j_m$
is zero.}
\label{bf}
\end{figure}
	
At half-filling
$z=0$ and, therefore, the current and the curvature of levels are,
up to order $t$, zero for
all the eigenstates of the model, leading to a zero value of $D(T)$.
This result agrees with that of Fujimoto and Kawakami
\cite{Kawakami98} if we take $U=\infty$ in their result.

If, for electronic densities equal to one, $t^2/U$ 
corrections are included, the flux dependence
of the energy levels is due to the rapidities 
$R_{c,\gamma,j}^{\infty}$ only. The flux dependence of
$R_{c,\gamma,j}^{\infty}$ controls
the behavior of $B^{c}_m$ with $\phi/L$. This factor is the only
responsible 
for the flux dependence of the energy eigenstates,
as can be seen from Eq. (\ref{eoveru}). 

To simplify our  study, we consider  only the $c,1$ string excitations.
From Eqs. (\ref{nbaixo}) and (\ref{nc})
it is clear that the creation of $N_{c,1}$ 
charge-string excitations such that
$N_{\uparrow}=N_{\downarrow}=$Const
introduces the constraints

\begin{eqnarray}
	\Delta N_{c}= N_{c}-N_{c}^{gs}=-2\Delta N_{c,1}=-2
N_{c,1}\nonumber\\
	\Delta N_{s,0}= N_{s,0}-N_{s,0}^{gs}=-\Delta
N_{c,1}=-N_{c,1}\nonumber\\
	N_{c}^{gs}=N\,, \hspace{1cm}N_{s,0}^{gs}=N_{\downarrow}\,,
\label{constr}
\end{eqnarray}
where $\Delta N_{\alpha,\gamma}$ represents the variation of the numbers
$N_{\alpha,\gamma}$ relatively to the ground state 
(only spinon excitations
have been considered) and $N_{\alpha,0}^{gs}$ is the number
of the occupied quantum numbers $I^{\alpha,0}$
in the ground state.

From Eq.(\ref{eoveru}) we see that the spin degrees of freedom do not
contribute to the flux dependence of the energy levels 
since they do not
depend in any way on $\phi/L$, up to order $t^2/U$. 
All the flux dependence of the energy levels is in the term

\begin{equation}
B=\sum_{j=1}^{N_{c,1}}
\frac {1}{1+(R_{c,1,j}^{\infty})^2}\,,
\label{bmc1}
\end{equation}
with $B$ depending on the eigenstate $m$ considered.

As we have stressed, up to order $t^2/U$ the charge gapped 
excitations are mathematically
equivalent (apart from the gap relative to the $c,0$-excitations) to an
isotropic 
Heisenberg model in a lattice of size $(L-N_c)$. Combining this
similarity with
Eq. (\ref{constr}), the sum $B$ reads, for large $N_{c,1}$, 
$B=N_{c,1}\ln 2$, at $\phi=0$.
At finite $\phi$, the sum $B$ depends on $\phi$, but presents a maximum
for
$\phi=0$, as can be seen from Figure \ref{bf}. If we take, as an example, 
the case
$L-N_c=2$ it simple to see that Eqs. (\ref{tak2inf}) and (\ref{bmc1}) do
reproduce the  flux dependence of the energy of a two-site Heisenberg chain.

%%%%%%%%%%%%%%%%%%%%%%%%%%%%%%%%%%%%%%%%%%%%%%%%%%%%%%%%%%%%%%%%%%%%%%%%%%%%%%
\section{Algebraic solution of the Hubbard chain with twisted
boundary conditions}
\label{algeb}

While the knowledge of the low-energy
eigenstates is enough if we are interested in the
low-energy or temperature behavior of the Hubbard model,
when calculating
the conductivity of the Hubbard model at half filling,
states with energies of the order of $U$ become relevant.
In this section, we obtain all eigenstates and eigenvalues of the Hubbard
model with a magnetic flux, when $t/U$ is a small parameter taking an
alternative approach to that of the BA method 
\cite{Lieb,Nuno97,Woynarovich82}, used in  the previous section.
 
%%%%%%%%%%%%%%%%%%%%%%%%%%%%%%%%%%%%%%%%%%%%%%%%%%%%%%%%%%%%%%%%%%%%%%%%%%
%                        The Hubbard model                               %
%%%%%%%%%%%%%%%%%%%%%%%%%%%%%%%%%%%%%%%%%%%%%%%%%%%%%%%%%%%%%%%%%%%%%%%%%%
The eigenvalues of the Hubbard model, when $t=0$,
are given by $E_{N_d}=N_d.U$ and all states
with a given number $N_d$ of doubly occupied sites are degenerate.
Using the identity $c_{i\sigma}=c_{i\sigma}[(1-n_{i\sigma})+n_{i\sigma}]$
\cite{harr},
where $c_{i,\sigma}$ is the fermion annihilation operator on site
$i$ with spin $\sigma$ and $n_{i,\sigma}=c_{i,\sigma}^\dagger
c_{i,\sigma}$,
the Hubbard model, Eq. (\ref{hamilt}), with $t\neq 0$ can be rewritten as
\cite{harr}
\begin{equation}
       \hat{H} = \hat{T}_h+\hat{T}_d+
       (\hat{T}_{hd}+\hat{T}_{dh})
       +U \sum_i n_{i,\uparrow} n_{i,\downarrow}
       \label{hamiltonian}
\end{equation}
where
\begin{equation}
      \hat{T}_h=-t \sum_{\langle ij \rangle \sigma} 
      (1-n_{i\bar{\sigma}})c^\dagger_{i\sigma} c_{j\sigma}
      (1-n_{j\bar{\sigma}}),
\end{equation}
\begin{equation}
      \hat{T}_d=-t \sum_{\langle ij \rangle \sigma} 
      n_{i\bar{\sigma}} c^\dagger_{i\sigma}
      c_{j\sigma} n_{j\bar{\sigma}},
\end{equation}
\begin{equation}
      \hat{T}_{dh}=-t \sum_{\langle ij \rangle \sigma} 
      n_{i\bar{\sigma}} c^\dagger_{i\sigma}
      c_{j\sigma} (1-n_{j\bar{\sigma}})
      ,
\end{equation}
with $\hat{T}_{dh}=\hat{T}_{hd}^\dagger$ and $\bar{\sigma}=-\sigma$.
Here, $\hat{T}_d$ and $\hat{T}_h$ describe the
movement of double occupations and
holes, respectively, but these terms and
the Hubbard onsite interaction do not
change the number of double occupations of a state.
An applied magnetic flux is easily introduced as a twisted boundary
condition.
In the limit of $U/t \rightarrow \infty$, the first correction to
the $t=0$ spectrum is obtained by diagonalizing the model given by
Eq. (\ref{hamiltonian})
without the mixing term $\hat{T}_{hd}+\hat{T}_{dh}$
within each subspace $N_d$. Such
diagonalization is described below.

%%%%%%%%%%%%%%%%%%%%%%%%%%%%%%%%%%%%%%%%%%%%%%%%%%%%%%%%%%%%%%%%%%%%%%%%%%
%            General case: $N_h$ holes, $N_d$ doubly occupancies         %
%%%%%%%%%%%%%%%%%%%%%%%%%%%%%%%%%%%%%%%%%%%%%%%%%%%%%%%%%%%%%%%%%%%%%%%%%%
\subsection{Exact diagonalization using slave-fermions: $U\gg t$ results}

Let us consider a ring of $L$ sites (labeled clockwise) with $L$ even and
a general state with $N_h$ holes and $N_d$ double occupancies
($N_s$ singly occupied sites) with positions defined by the ordered sets
$\{h\}$
and $\{d\}$ and with the spin configuration of the singly occupied sites
given
by the set $\{\sigma\}$:
\begin{eqnarray}
        \{h\} &=& \{h_1,\dots ,h_{N_h}\}; \nonumber \\
        \{d\} &=& \{d_1,\dots , d_{N_d}\}; \nonumber \\
        \{\sigma\} &=& \{\sigma_1, \dots, \sigma_{L-N_h-N_d}\} \nonumber
\end{eqnarray}
\begin{equation}
        \vert \{h\};\{d\};  \{\sigma\} \rangle =
        \prod^{N_d}_{i=1} c^\dagger_{d_i \uparrow}
        c^\dagger_{d_i \downarrow}
        \prod^{N_s}_{j=1} c^\dagger_{b_j \sigma_j} \vert 0 \rangle
        \label{general}
\end{equation}
with $b_j=j+n_j(\{h\},\{d\})$ where $n_j(\{h\},\{d\})$ is the number of
holes plus double occupancies to the left of site $j$.
The Hamiltonian applied to this state exchanges a hole or a
double occupancy with a spin, without
changing the spin configuration $\{ \sigma \}$ except for
hoppings at the boundaries where the final spin configuration
is a cyclic permutation of the original one.

We now introduce the slave-fermion representation
\cite{Dias92,hubb,slave} for the fermion operators,
${c}_{i,\sigma}=e_i^{\dagger} S_{i\sigma}+\sigma S^{\dagger}_{i,-\sigma}
d_i$, where $e_i^{\dagger}$ and $d_i^{\dagger}$
are operators satisfying fermionic commutation relations
($\{e_i^{\dagger},d_i^{\dagger}\}=0$)
and $S_{i\sigma}$ is a spinon operator satisfying
bosonic commutation relations
with the restriction $e_i^{\dagger}e_i+ d_i^{\dagger}d_i +\sum_{i\sigma}
S_{i\sigma}^{\dagger} S_{i\sigma}=1$.
Let us define the ordered set
$\{a\}=\{h\} \cup \{d\}=\{a_1,\dots ,a_{N_h+N_d}\}$ and a
set of fictitious spins $\{\nu\}=\{\nu_1,\dots ,\nu_{N_h+N_d}\}$
which indicates whether we have
a hole or a double occupancy at $a_i$, that is,
$$
        C^\dagger_{a_i\uparrow}= e^\dagger_{a_i} ; \quad
        C^\dagger_{a_i\downarrow}= d^\dagger_{a_i}.
$$
These operators obey fermionic commutation relations.
Our general state in the slave-fermion representation, using
the previous definition, becomes
\begin{eqnarray}
        \vert \{a\};\{\nu\};  \{\sigma\} \rangle &=&
        \prod^{N_s}_{j=1} S^\dagger_{b_j \sigma_j}
        \prod^{N_h+N_d}_{k=1} C^\dagger_{a_k \nu_k}
        \vert 0 \rangle_{sf} \nonumber \\
        &\times & (-1)^{\sum_{i=1}^{N_s} (b_i-1)} \label{state}
\end{eqnarray}
and the Hamiltonian is
\begin{equation}
        \hat{H}-N_d\cdot U= \sum_{i=1 \atop \sigma \nu}^L
        t_{i\nu} S_{i+1,\sigma}^\dagger S_{i,\sigma}
        C_{i,\nu}^\dagger
        C_{i+1,\nu} +\mbox{h.c.}
        \label{eq:hamiltonian1}
\end{equation}
with $t_{i\nu}=t_{\nu}=\nu \cdot t$ for $i\neq L$ and
$t_{L\nu}=e^{i\nu \cdot \phi} t_{\nu}$. A $\nu$-spin dependence
of the hopping integral just reflects that the hole band is shifted
by $\pi$ relatively to the electron band. Note that with this
representation,
the boundary conditions also become $\nu$-spin dependent, since
a hopping of a hole from site L to 1 implies an opposite hopping of an
electron
from 1 to L.
As we shall see, this will lead to an anomalous flux dependence of the
$U\gg t$
eigenstates.

Let us define 
\begin{eqnarray} 
      \lefteqn{\vert a_1',\dots ,a_{N_h+N_d}'; \nu_1', \dots,
      \nu_{L-N_h-N_d}' \rangle= }
      \quad \quad && \\ 
      &&  C_{i+1,\nu}^\dagger C_{i,\nu}
      \vert a_1,\dots ,a_{N_h+N_d}; \nu_1, \dots,
      \nu_{L-N_h-N_d} \rangle \nonumber 
\end{eqnarray} 
In order to simplify this Hamiltonian, we construct Bloch states
in the cyclic permutations of the spin configuration $\{\sigma\}$,
\begin{eqnarray}
      \lefteqn{\vert \{h\};\{d\}; {\alpha_s},q_s \ra = }
      \quad \quad && \\
      && \quad \quad {1\over \sqrt{r_{\alpha_s}}}
      \sum_{m=0}^{r_{\alpha_s}-1} e^{iq_sm}
      \vert \{h\};\{d\}; \sigma_{1-m},\cdots ,\sigma_{L-2-m}\ra .
      \nonumber
\end{eqnarray}
where $q_s=n(2\pi/r_{\alpha_s})$ with $n=0, \cdots , r_{\alpha_s}-1$ and
$r_{\alpha_s}$ is the period of the spin configuration.
The ${\alpha_s}$ index labels the different (not obtainable from any other
by cyclic permutations) spin configurations with period $r_{\alpha_s}$.
Then
\begin{eqnarray}
        \lefteqn{\sum_{\sigma}
        S_{i,\sigma}^\dagger S_{i+1,\sigma}
        C_{i+1,\nu}^\dagger
        C_{i,\nu}
        \vert \{a\};\{\nu\}; {\alpha_s}, q_s \rangle =}
        && \\
        &&  \hspace{2cm} = \left\{
        \begin{array}{llc}
                (-1) &\vert \{a'\};
                \{\nu '\}; {\alpha_s}, q_s \ra,
                & i\neq L\\
                (-1) e^{iq_s}
                & \vert \{a'\};
                \{\nu '\}; {\alpha_s}, q_s \ra,
                & i= L  \nonumber
        \end{array}
        \right.
\end{eqnarray}
For the state given in Eq. (\ref{state}), the successive application of
$\hat{T}_h$ and $\hat{T}_d$ does not change the sequence of
these fictitious spins $\{\nu\}$,
but permutates them cyclicly as in the case of the real spins
$\{\sigma\}$.
Given a $\{{\alpha_s},q_s\}$ subspace, the Hamiltonian can be written as:
\begin{eqnarray}
        \hat{H}(q_s) &=& \sum_{i=1 , \nu}^L
        [t_{i\nu}
        (1-N_{i+1,-\nu}) C_{i+1,\nu}^\dagger
        C_{i,\nu}(1-N_{i,-\nu})
        \nonumber \\
        && \quad \quad +\mbox{h.c.}]+ N_d\cdot U
\end{eqnarray}
with $t_{i\nu}=-t_{\nu}$, $i\neq L$,
$t_{L\nu}=- e^{iq_s} e^{-i\nu \cdot \phi} t_{\nu}$,
and $N_{i,\nu}=C_{i,\nu}^\dagger C_{i,\nu}$.

\begin{figure}[htb] 
\begin{center} 
\leavevmode 
\hbox{% 
\epsfxsize=3.0in \epsfbox{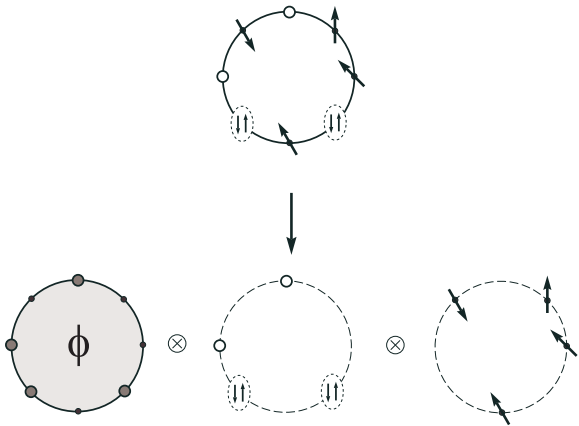}} 
\end{center} 
\caption{The high energy eigenstates of our model are mapped 
         onto a system of free fermions on a chain threaded by 
         a fictitious flux $\phi=(q_s-q_c)\phi_0/\pi$ and 
         two squeezed chains, one of holes and double occupancies and 
         the other of spins.} 
\label{fig:2decouple} 
\end{figure}

Let us consider first the zero flux case, $\phi=0$.
In that case, the model above is the
one-band ($U=\infty$) Hubbard model with twisted boundary
conditions, except for the spin dependence of the hopping integral.
In order to make $t_{i\downarrow}=t_{i\uparrow}$, we make the
following gauge transformation: 
$$
        C_{j,\uparrow}^\dagger
        \rightarrow C_{j,\uparrow}^\dagger e^{-i\pi j}.
$$
Then $t_{i\nu}\rightarrow t$, $i\neq L$ and $t_{L\nu}= e^{iq_s} t$.
Now, the gauge transformation
$$
       C_{j,\nu}^\dagger
       \rightarrow C_{j,\nu}^\dagger e^{i{q_s\over L} j}
$$ 
leads to
$t_{j\nu}\rightarrow te^{i{q_s\over L}}$, $\forall j,\nu$.
The solution of this model is known (this is the $U=\infty$
Hubbard chain enclosing a magnetic flux $\phi_o q_s/\pi$),
but now our up and down spins
indicate whether the charge carrier is a hole or a double occupancy. The
spin-charge factorization which we know characterizes the eigenfunctions
of the $U=\infty$ Hubbard model
(we refer the readers to Ref. \cite{Dias92} for a detailed solution of
this
model)
translates to a further decoupling of the part of the
wavefunction describing the charge degrees of freedom and consequently,
the
eigenstates are factorized in the form:
$ \vert ${\it charge carriers positions}$\ra \otimes
\vert ${\it hole/double occupancy configuration}$\ra \otimes
\vert ${\it spin configuration}$\ra $
(see Fig.~\ref{fig:2decouple}).
That is, the eigenstates of our model can be mapped
onto a system of fermions on a chain with $L$ sites and
two squeezed chains, one of holes and double occupancies and
the other of spins.
The eigenstates will be of the form 
$$
      \vert \{k\}\ra \otimes \vert \alpha_c , q_c
       \ra \otimes \vert {\alpha_s}, q_s \ra 
$$
where $q_c$ is the momentum of the Bloch state in the cyclic permutations
of the $\{\nu\}$ configuration, $q_c=n(2\pi/r_{\alpha_c})$
with $n=0, \cdots , r_{\alpha_c}-1$ and
$r_{\alpha_c}$ is the period of the holes/double occupancies
configuration.
The charge carriers will be free fermions in a chain threaded by a
fictitious magnetic flux $\phi=(q_s-q_c)\phi_0/\pi$
generated by {\it both} the spin and hole/double occupancy
configurations and the eigenvalues of
$\hat{H}(q_s)-N_d\cdot U$ are given by
\begin{equation}
        E(k_1,\cdots ,k_{N_h+N_d})=2t \sum_{i=1}^{N_h+N_d}
        \cos \left(k_i-{q_s-q_c \over L}\right),
        \label{eq:eigen}
\end{equation}
with $k_i=(2\pi/L) n_i$, $n_i=0,\cdots,L-1$ \cite{comment}.
The fact that the hole/double occupancy configuration also generates
a fictitious flux implies that the lowest eigenvalue of a $N_d\neq 0$
subspace may have an even weaker dependence on an applied flux
than that of the ground state energy \cite{scho}.

%%%%%%%%%%%%%%%%%%% 
%    figure       % 
%%%%%%%%%%%%%%%%%%% 

In order to solve the $\phi \neq 0$ case,
we first repeat the gauge transformations 
$C_{j,\uparrow}^\dagger 
\rightarrow C_{j,\uparrow}^\dagger e^{-i\pi j}$
and
$C_{j,\nu}^\dagger 
\rightarrow C_{j,\nu}^\dagger e^{i{q_s\over L} j}$ so that
$t_{j\nu}\rightarrow te^{i{q_s\over L}}$, $\forall j\neq L$
and $t_{L\nu}\rightarrow t e^{i{q_s\over L}}e^{-i\nu \cdot \phi}$. 
Now, we define a operator $\hat{Q}$ such that
\begin{equation}
\hat{Q} \vert \nu_1, \dots, \nu_{L-2}\rangle =
\vert \nu_2, \dots, \nu_{L-2}, \nu_1 \rangle\,,
\end{equation}
that is, it does a circular permutation of the spin configuration.
For the $\phi \neq 0$ case, given a configuration of fictitious spins
$\{\nu\}$,
it is obvious the eigenstates will be found in the subspace spanned
by $\hat{Q}^n \{\nu\}$, $n=0,\cdots , r_{\alpha_c}-1$.
The hoppings across the boundary do a cyclic permutation with a  factor
$e^{i\nu \cdot \phi}$.
We wish to construct now the state such that
\begin{equation}
        \hat{Q}_{\nu_1} \left( \sum_{i=0}^{r_{\alpha_c}-1} a_i
        \hat{Q}^i \vert \{\nu\}\ra \right) = e^{i\phi'}
        \left( \sum_{i=0}^{r_{\alpha_c}-1}
        a_i \hat{Q}^i\vert \{\nu\}\ra \right),
\end{equation}
where $\hat{Q}_{\nu_1}=\hat{Q} e^{i \phi}$ if $\nu_1=\uparrow$ and
$\hat{Q}_{\nu_1}=\hat{Q} e^{-i \phi}$ if $\nu_1=\downarrow$. $\phi'$ will
be
the effective flux felt by the noninteracting fermions.
That is, we want to construct the equivalent to the Bloch states
in the cyclic permutations of $\{\sigma\}$ for the $\phi=0$ case.

This problem is equivalent to solving a one-particle tight-binding model
for a chain of $r_{\alpha_c}$ sites with
hopping constant $t_{j}= t e^{i\nu_j\cdot \phi}$ with the 
correspondence $\vert i \rangle =\hat{Q}^{i-1} \vert \{\nu \} \rangle$.
The total flux through this tight-binding chain is 
\begin{equation} 
          \phi_1= r_{\alpha_c} {N_h-N_d \over N_h+N_d} \phi
\end{equation} 
The solution is obtained after a gauge transformation to make
$t_{j} \rightarrow e^{i\phi_1 /r_{\alpha_c}} t$. The gauge transformation 
depends on the $\nu$-spin configuration, but the tight binding eigenvalues 
only depend on the total flux. The eigenstates will be Bloch states $\vert
\alpha_c, q_c \ra$ (in the cyclic permutations) in the way defined for
$\phi=0$ with $q_c=n(2\pi/r_{\alpha_c})$ 
with $n=0, \cdots , r_{\alpha_c}-1$ and eigenvalues given by
\begin{equation} 
          \epsilon(q_c)=2t \cos \left( q_c-{\phi_1 \over r_{\alpha_c}}
\right)
\end{equation}  
This implies 
\begin{equation} 
        \hat{Q}_{\nu_1} \vert \alpha_c, q_c \ra = 
        e^{-i \left( q_c-{\phi_1 \over r_{\alpha_c}} \right)} \vert
\alpha_c, q_c \ra
\end{equation}  
and therefore, repeating the $\phi=0$ steps,
the Hamiltonian can diagonalized and the eigenvalues of
$\hat{H}(q_s)-N_d\cdot U$ 
are given by
\begin{eqnarray}
        E(k_1,\cdots ,k_{N_h+N_d}) &=& 2t \sum_{i=1}^{N_h+N_d}
        \cos \left[k_i-{q_s-q_c \over L}  \right.
        \quad \quad \\
        && \hspace{2cm}
        \left. -{1 \over L} {N_h-N_d \over N_d+N_h} \phi \right],
\nonumber
\end{eqnarray}
with $k_i=(2\pi/L) n_i$, $n_1=0,\cdots,L-1$.
%%%%%%%%%%%%%%%%%%%%%%%%%%%%%%%%%%%%%%%%%%%%%%%%%%%%%%%%%%%%%%%%%%%%%%%%%%
%                         Phase shifts                                   %
%%%%%%%%%%%%%%%%%%%%%%%%%%%%%%%%%%%%%%%%%%%%%%%%%%%%%%%%%%%%%%%%%%%%%%%%%%
Summing up, we have derived an
effective Hamiltonian for our spinless and chargeless particles
\begin{equation} 
     \hat{H}(q_c,q_s) = t  e^{-i {\Phi\over L}} 
     \sum_{i} \beta_{i}^\dagger
     \beta_{i+1} + \mbox{h.c.}.
\end{equation}  
with
\begin{equation} 
     \Phi=\Phi^{stat}+\Phi^{eff}\,,
\label{rphieff}
\end{equation}
and $\Phi^ {stat}$ and $\Phi^{eff}$ given by     
\begin{equation}
\Phi^{stat}=q_s-q_c\,,
\label{rpstat}
\end{equation}
and
\begin{equation}
\Phi^{eff}={N_h-N_d \over N_d+N_h} \phi\,.
\label{rpeff}
\end{equation}
The Hamiltonian describes free fermions in a chain threaded
by a fictitious magnetic flux $\phi_o (q_s-q_c)/\pi$
(see Fig. \ref{fig:2decouple}).
The lowest energy state is obtained for $q_s-q_c=0$ and $q_s-q_c=\pi$
for an odd and even number of holes plus double occupancies respectively
in the case of zero external flux.
The flux given by Eq. (\ref{rphieff}) has exactly the same form
as obtained previously. The $\Phi^{stat}$ term corresponds to the sums
over the $I^{s,\gamma}_j$ and $I^{c,\gamma}_j$ numbers in Eq.
(\ref{tak1inftot}), and the term $\Phi^{eff}$ is the same as in 
Eq. (\ref{fluxterm2}). It quite clear from this approach that the presence
of doubly occupied sites is responsible for the renormalization
of the flux felt by the holons.

%%%%%%%%%%%%%%%%%%%%%%%%%%%%%%%%%%%%%%%%%%%%%%%%%%%%%%%%%%%%%%%%%%%%%%%%%%
%                         canonical transformation                       %
%%%%%%%%%%%%%%%%%%%%%%%%%%%%%%%%%%%%%%%%%%%%%%%%%%%%%%%%%%%%%%%%%%%%%%%%%%
\subsection{Transport of charge}

In the limit $t \ll U$, an effective Hamiltonian to second order
in $t$ can be obtained for each $N_d$ subspace by a canonical
transformation
$\hat{H}_{eff}=e^{iS} \hat{H} e^{-iS}$ \cite{harr} leading
to the additional term $(\hat{T}_{hd}\hat{T}_{dh}-
\hat{T}_{dh}\hat{T}_{hd})/U$, ($N_d\neq 0$).
This term will give a correction of order $t^2/U$ 
and lift the degeneracy of states $\vert \{k\}; {\alpha_c}, q_c;
{\alpha_s}, q_s \ra $ in what concerns the spin and
charge configurations $\vert {\alpha_c}, q_c; {\alpha_s}, q_s \ra $
as in the case of $N_d=0$ subspace where a similar term leads to
a Heisenberg interaction plus a three-site term \cite{Ogata90,Dias92}.

%%%%%%%%%%%%%%%%%%%%%%%%%%%%%%%%%%%%%%%%%%%%%%%%%%%%%%%%%%%%%%%%%%%%%%%%%% 
%                            Hubbard gap                                 % 
%%%%%%%%%%%%%%%%%%%%%%%%%%%%%%%%%%%%%%%%%%%%%%%%%%%%%%%%%%%%%%%%%%%%%%%%%% 
The Hubbard gap, that is, the energy difference between the highest 
eigenvalue with $N_d$ double occupancies and the lowest eigenvalue with 
$N_d+1$ double occupancies for a system with N electrons, 
is easily obtained from Eq. (\ref{eq:eigen}), 
\begin{equation} 
        \Delta(L,U)=U-4t \cos \left({2\pi \over L} {N_h+N_d \over 2} 
        +\delta \right),
       \label{rgap}	 
\end{equation} 
with $\delta=0$ for $N_h+N_d$ odd and $\delta=\pi/L$ for $N_h+N_d$ even. 
For a half-filled system, $N_h=N_d$ and the gap grows with $N_d$.  Eq.
(\ref{rgap}) is the same as Eq. (\ref{gaplu}), if $t^2/U$ corrections
are neglected.
 
%%%%%%%%%%%%%%%%%%%%%%%%%%%%%%%%%%%%%%%%%%%%%%%%%%%%%%%%%%%%%%%%%%%%%%%%%% 
%                         charge stiffness                               % 
%%%%%%%%%%%%%%%%%%%%%%%%%%%%%%%%%%%%%%%%%%%%%%%%%%%%%%%%%%%%%%%%%%%%%%%%%% 
The Hubbard Hamiltonian can be written as a sum of a diagonal term 
in the number of double occupancies and a mixing term which has little 
effect in the strong coupling limit \cite{Dias92}. 
Similarly, for the current operator, 
\begin{equation}\label{current} 
      J=it\sum_{i,\sigma} (c_{i,\sigma}^\dagger c_{i+1,\sigma} 
      - c_{i+1,\sigma}^\dagger c_{i,\sigma}) 
\end{equation} 
we can write $J=J_o+J_1$ where 
$J_o$ commutes with $H(\phi=0)$ and $J_1$ is the mixing term. These two
operators 
in the slave-fermion representation are given by 
\begin{equation}\label{jo} 
      J_o=it \sum_{i ,\sigma,\nu} 
      (S_{i+1,\sigma}^\dagger S_{i,\sigma} 
      C_{i,\nu}^\dagger C_{i+1,\nu} 
      -S_{i,\sigma}^\dagger S_{i+1,\sigma} 
      C_{i+1,\nu}^\dagger C_{i,\nu}) 
\end{equation} 
and 
\begin{equation}\label{j1} 
      J_1=i2t \sum_{i} 
      ({\mathcal C}_{i+1,i}^\dagger {\mathcal S}_{i+1,i} 
      -{\mathcal C}_{i+1,i} {\mathcal S}_{i+1,i}^\dagger) 
\end{equation} 
where 
\begin{equation} 
      {\mathcal C}_{i+1,i}^\dagger ={1 \over \sqrt{2}} 
      (C_{i+1,\uparrow}^\dagger C_{i,\downarrow}^\dagger 
      -C_{i+1,\downarrow}^\dagger C_{i,\uparrow}^\dagger) 
      \nonumber 
\end{equation} 
\begin{equation} 
      {\mathcal S}_{i+1,i}^\dagger ={1 \over \sqrt{2}} 
      (S_{i+1,\uparrow}^\dagger S_{i,\downarrow}^\dagger 
      -S_{i+1,\downarrow}^\dagger S_{i,\uparrow}^\dagger) 
      \nonumber 
\end{equation} 
Note that in $J_o$, no $\nu$-spin dependence appears in the hopping
integral
because the term in this current operator 
that hops a double occupancy from $i$ to $i+1$
has a minus sign while the term that hops a hole from $i$ to $i+1$
does not. The $J_o$ current operator is now simply the double
occupancy current minus the hole current.
 
One should note that when determining the optical conductivity, 
one has to calculate matrix elements of the current operator 
between states with $N_d$ and $N_d+1$ double occupancies 
in order to obtain the upper Hubbard band part of the optical 
conductivity. Obviously, we can replace $J$ by $J_1$ in this case. 
The low frequency region is given by matrix elements of the current 
operator between states with the same number of double occupancies 
and in this case, we replace $J$ by $J_o$. 

The expectation value of the current operator for our zero flux
eigenstates
can be easily obtained if we rewrite the current operator as
\begin{equation}\label{jo1}  
      J_o=i \sum_{i,\sigma,\nu} (2 {\mathcal C}^z_i ) \cdot t_{i\nu}
      S_{i+1,\sigma}^\dagger S_{i,\sigma}  
      C_{i,\nu}^\dagger C_{i+1,\nu} +h.c.
\end{equation}  
where ${\mathcal C}^z_i= (C_{i,\uparrow}^\dagger C_{i,\uparrow} 
-C_{i,\downarrow}^\dagger C_{i,\downarrow})/2$ and with $t_{i\nu}$
given as in Eq. \ref{eq:hamiltonian1}. This term just means 
that a hole current implies an electron current in the opposite direction,
but 
a double occupancy current is an electron current. The second term is of 
the same form as that in Eq. (\ref{eq:hamiltonian1}) and applied to an
eigenstate,
it would multiply it by a phase if the ${\mathcal C}^z_i$ term was not
present. 
One should note that if we consider the squeezed $\nu$-spin chain,
$ \langle \{ \nu\} \vert {\mathcal C}^z_m  \vert \{ \nu\} \ra$ will give
the
value of the spin in site $m$, but 
$\langle \alpha_c , q_c \vert {\mathcal C}^z_m  \vert \alpha_c , q_c  \ra$
is independent of $m$ since the state $\vert \alpha_c , q_c  \ra$
is invariant under a $\nu$-spin translation. We can therefore replace it
by a sum 
over all sites in the reduced $\nu$-spin chain, divided by the total
number of sites
of this squeezed chain, $N_d+N_h$. And obviously,
\begin{equation}  
       {1 \over N_d+N_h} \langle \alpha_c , q_c \vert 
       \sum_{m }2 {\mathcal C}^z_m 
       \vert \alpha_c , q_c  \ra ={N_h-N_d \over N_h+N_d}
\end{equation}  
Clearly, the average value of $J_1$ is zero and so, using the above
expression
for $J_o$, we obtain       
\begin{eqnarray} 
       \langle \{k\};\alpha_c , q_c ;\alpha_s , q_s \vert J
       \vert \{k\}; \alpha_c , q_c ;{\alpha_s}, q_s \ra 
       &=& \\ 
       && \hspace{-4cm} 
        -2t {N_h-N_d \over N_h+N_d} \sum_{i=1}^{N_h+N_d} 
       \sin \left[k_i-{q_s-q_c \over L} \right] \nonumber\,. 
\end{eqnarray}
The equation derived above for the mean value of the current operator
is the same as Eq. (\ref{currz}), derived from the BA, it leads to
$D(T)=0$, at half filling, and to $D(T)\neq 0$ otherwise.  

This result and the ones described previously
suggest a simple picture for the $U\gg t$ Hubbard model,
that of a tight-binding model  of $N_d+N_h$ fermions with an
average charge $(N_h-N_d)/(N_h+N_d)$ moving in a averaged spin
background $\langle S^z \rangle =1/2 \cdot
(N_\uparrow-N_\downarrow)/((N_\uparrow+N_\downarrow)$.

%%%%%%%%%%%%%%%%%%%%%%%%%%%%%%%%%%%%%%%%%%%%%%%%%%%%%%%%%%%%%%%%%%%%%%%%%%%%
\section{Comparison with an ideal semi conductor}
\label{semi}

\begin{figure}[f]
\epsfxsize=6.5 cm 
\epsfysize=8.0 cm
\centerline{\epsffile{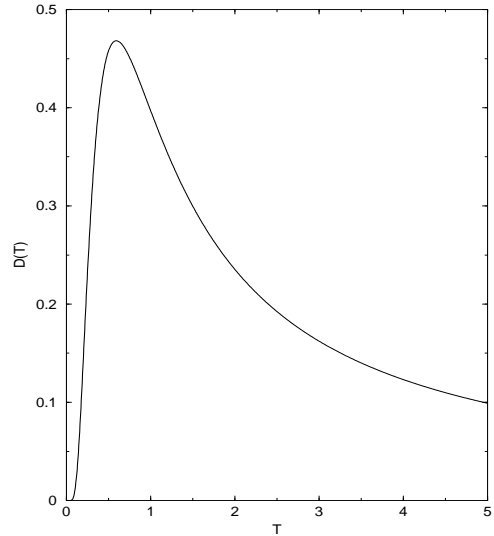}}
\caption{Charge stiffness for the dimerized model, at half-filling. The
parameters
are $a=1$, $t_1=1$, $t_2=2$. The temperature axis is scaled by the gap
$\Delta$.}
\label{dtfree}
\end{figure}

The simplest one-dimensional model-Hamiltonian for an ideal semi-conductor
is the dimerized tight-binding model, which, after diagonalization, reads
(the spin index is omitted)\cite{Gebhard}
\begin{equation}
	H=\sum_{k,\beta}\beta E(k)a^{\dag}_{k,\beta}a_{k,\beta}\,,
\end{equation}
with
\begin{equation}
E(k)=\sqrt{t_1^2+t_2^2+2t_1t_2\cos(2ka)}\,,
\end{equation}
and $t_{1,2}$ are the hopping integrals, $a$ the lattice spacing, $\beta$
the
band index,
and $k$ the momentum, belonging to the 
Brillouin zone  $[-\pi/(2a),\pi/(2a)]$.  The zero-temperature charge
stiffness
is easily computed, and is given by
\begin{equation}
D(0)=\frac {4t_1t_2}{\pi} \frac{\sin
(k_Fa)\cos(k_Fa)}{[(t_1+t_2)^2-4t_1t_2
\sin^2(k_Fa)]^{1/2}}\,,
\end{equation}
with $k_F$ the Fermi momentum. At zero temperature,
$D(0)$  is zero at half filling, the contributions to $D(0)$ being only
from the lower band $\beta=-1$. At finite temperature there are
contributions
to $D(T)$ from particles in both bands. However, and at odds with the
Hubbard model,
finite occupancies of the upper band ($\beta=1$) does not renormalize the
flux felt by the holes that remain in the lower band. For the
dimerized model we obtain
a finite charge stiffness at half filling, given by
\begin{equation}
D(T)=\frac{2}{\pi}\int_{-\pi/(2a)}^{\pi/(2a)}dkf(k,T)d(k)\,,
\end{equation}
with $d(k)$ given by
\begin{equation}
d(k)=-\frac{E^4(k)-(t^2_2-t^2_1)^2}{E^3(k)}\,,
\end{equation}
and $f(k,T)=(1+\exp(E(k)/T))^{-1}$ (the chemical potential is zero).
We plot $D(T)$ in Figure \ref{dtfree}. From this Figure we see that the
maximum
of $D(T)$ is controlled by the energy gap $\Delta=2\vert t_1-t_2\vert$.
In the Hubbard model had not
the effective flux felt by the holons been
renormalized to zero
at half filling, and we would have obtained for that model a physical
behavior 
similar to that in Figure \ref{dtfree}. At $T=0$ and $n=1$, we can still
view 
the Hubbard  model as a {\it band insulator}, since the Mott-Hubbard transition
has its origin in the gap between the bands of $c,0$ and of the 
$c,\gamma$ excitations.

%%%%%%%%%%%%%%%%%%%%%%%%%%%%%%%%%%%%%%%%%%%%%%%%%%%%%%%%%%%%%%%%%%%%%%%%%%%%%%%%%%%%%%
\section{conclusions}
\label{conclu}

In this paper we have studied the behavior of the finite-temperature 
charge stiffness for Hubbard rings of finite-size $L$. 

For  $U=\infty$ and $n<1$, we have seen that $D(T,L)$ is finite, 
decreasing as the temperature increases. In finite rings there is
a difference between free spinless fermions and the
$c,0$ excitations of the Hubbard model, due to the
statistical flux $\phi^{stat}_s$. This statistical flux
is also responsible for the sensitivity of $D_2(T,L)$
-- the contribution to $D(T,L)$ from the square of the matrix 
elements of the current operator -- to the
inclusion of all the spin excitations. At high
temperatures we have seen that this sensitivity vanishes. We 
have also verified that $D(T,L)$, as calculated from the
curvature of levels $D_m$, is not very sensitive to the
inclusion of only some of the spin excitations, even at low temperatures. 
At half-filling, the only available state, apart from the massive spin
degeneracy, is the ground state which leads to $D(T)=0$. Our results
agree with the statement of Zotos and Prelov\v{s}ek
\cite {Zotos96} that $D(T)$ should be finite if $D(0)$ is also finite.

For $U\gg t$, we have derived 
asymptotic equations for the mean value of the current
operator and for 
the curvature of levels valid for any band filling. This
derivation is specially important at half-filling, where 
some conjectures by Zotos and Prelov\v{s}ek
\cite {Zotos96} indicate that $D(T,\infty)$ should be zero. These
authors related this odd behavior with the integrability of the
model. We have shown, using two independent
methods, that if doubly occupied sites are allowed the
flux felt by the low-energy charge excitations  -- holons --
is renormalized to zero, at $n=1$. For this density, only the
gapped charge-excitations  feel the flux piercing the ring, but
the current transported by these excitations is zero, and therefore,
$D(T)=0$.  We stress that his
result does not contradict what was obtained by Fujimoto
and Kawakami  \cite{Kawakami98}, since they also obtain $D(T)=0$ in the 
limit we are considering.
Nevertheless, our analysis shows that if $D(T)=0$ is to be finite, for
$n=1$, the
prefactor of the exponential of the Mott-Hubbard gap must be, at least, of the
order of $t^2/U$, and not of the order one, as it is found by
Fujimoto
and Kawakami \cite{Kawakami98}. 
The physical interpretation
of this effect is quite transparent in our approach: the
higher energy excitations are renormalizing to zero (for $n=1$)
the flux felt by the low-energy excitations. In addition it is
also clear from our approach that the highest occupied band is always
completely filled and has a maximum at $\phi=0$. This effect,
together with the renormalization of the flux leads to a zero value
of the current, in any eigenstate of the model. Another
way of looking at our results is that the $U\gg t$ Hubbard model,
can be thought as  a tight-binding model of $N_d+N_h$ fermions with an
average charge $(N_h-N_d)/(N_h+N_d)$ moving in a averaged spin
background $\langle S^z \rangle =1/2 \cdot
(N_\uparrow-N_\downarrow)/((N_\uparrow+N_\downarrow)$.

If the system is not at half filling and
doubly occupied sites are allowed, the flux felt by the holons
is renormalized away from the usual $\phi/L$
value by the presence of doubly occupied sites.

%%%%%%%%%%%%%%%%%%%%%%%%%%%%%%%%%%%%%%%%%%%%%%%%%%%%%%%%%%%%%%%%%%
\section*{ACKNOWLEDGMENTS}

We thank Miguel Ara\'{u}jo, Dionys Baeriswyl, David Campbell,
Jo\~ao Lopes dos Santos,
and Lu\'{\i}s Miguel 
Martelo for stimulating discussions.
J.M.P.C. is grateful for the hospitality
and financial support of N.O.R.D.I.T.A. in Copenhagen.
This research was funded by the Portuguese
MCT PRAXIS XXI program under Grant No. 2/2.1/Fis/302/94.
%%%%%%%%%%%%%%%%%%%%%%%%%%%%%%%%%%%%%%%%%%%%%%%%%%%%%%%%%%%%%%%%%%%%%%%%
%%%%%%%%%%%%%%%%%%%%REFERENCES%%%%%%%%%%%%%%%%%%%%%%%%%%%%%%%%%%%%%%%%%%

\end{multicols}
\widetext
\newpage
%%%%%%%%%%%%%%%%%%%%%%%%%%%%%%%%TABLE 1%%%%%%%%%%%%%%%%%

\begin{table}
\begin{tabular}{cccc}
quantum numbers & type of excitations & gap& rapidity\\
\tableline
$I^{c}_j$ & holons & gapless & $k_j$\\
$I^{s,0}_j$ & spinons & gapless & $R_{s,0,j}$\\
$I^{c,\gamma >0}_j$ & charge strings & gapped&$R_{c,\gamma,j}$\\
$I^{s,\gamma >0}_j$ & spin stings & gapped ($H\ne 0$)& $R_{s,\gamma,j}$\\
\end{tabular}
\caption{Types of excitations associated with the quantum
numbers $I^{\alpha,\gamma}_j$, with $\alpha=c,s$. The
spin-string excitations are gapped only for
finite magnetic field $H$. Both the holon and spinon
excitations can be described by bosonization methods.
The smaller gap associated with the charge string
exciations is the Mott-Hubbard gap $\Delta_{MH}$.}
\label{class}
\end{table}
%%%%%%%%%%%%%%%%%%%%%%%%%%%end table%%%%%%%%%%%%%%%%%

%%%%%%%%%%%%%%%%%%%%%%%%%%%%%%%%TABLE 2%%%%%%%%%%%%%%%%%
\begin{table}
\begin{tabular}{ccccc}
state & $N_{c}$ & $N_{s,0}$ & $N_{c,\gamma >0}$& $N_{s,\gamma >0}$\\
\tableline
G. S. & $N$ & $N_{\downarrow}$ & 0 & 0\\
Ex.$_0$ &$N$ & $N_{\downarrow}$ & 0 & 0\\
Ex.$_{c,\gamma}$ &  $N-2\gamma N_{c,\gamma}$& 
$N_{\downarrow}-\gamma N_{c,\gamma}$&$N_{c,\gamma}$&0\\
Ex.$_{s,\gamma}$&$N$&
$N_{\downarrow}-(\gamma+1)N_{s,\gamma}$&0&$N_{s,\gamma}$\\
\end{tabular}
\caption{The numbers, $N_{c,\gamma}$, of the different kinds of
excitations,
in different classes of eigenstates.
The notation is as follows: G. S., Ex.$_0$, Ex.$_{c,\gamma}$, and
Ex.$_{s,\gamma}$ stand for the ground state, the low-energy eigenstates
(no-strings), the eigenstates with $N_{c,\gamma}$ charge strings 
of length $\gamma$, and
the eigenstates with $N_{s,\gamma}$ spin string of length $\gamma$,
respectively.}
\label{stateg}
\end{table}
%%%%%%%%%%%%%%%%%%%%%%%%%%%end table%%%%%%%%%%%%%%%%%
 
%%%%%%%%%%%%%%%%%%%%%%%%%%%%%%%%TABLE 3%%%%%%%%%%%%%%%%%
\begin{table}
\begin{tabular}{cccccccccc}
states & $(I_j^{c})^{min}$&$N_{c}$ 
	 & $(I_j^{s,0})^{min}$&$N_{s,0}$ 
	 & $(I_j^{s,1})^{min}$&$N_{s,1}$
	 & $(I_j^{s,2})^{min}$&$N_{s,2}$
	 & $\phi^{stat}_s$\\
\tableline
Ex.$_0$ & -11/2 & 6&-1&3& --&--&--&--&0     \\
Ex.$_{s,1}$ & -5(-6)& 6& -1&1&0&1&--&--&
	$-\frac{\pi}{36},0,\frac{\pi}{36}$\\
Ex.$_{s,2}$&  -11/2&6&--&--&--&--&0&1&0\\
\end{tabular}
\caption{The notation is as follows:  Ex.$_0$, Ex.$_{s,1}$, and
Ex.$_{s,2}$ stand for the states with no spin strings, the eigenstates
with 
one $\gamma=1$ spin-string, and the eigenstates with 
one $\gamma=2$ spin-string, respectively. 
The system size is $L=12$, the number of electrons
is $N_{\uparrow}=N_{\downarrow}=3$, and $U=\infty$. The number of occupied
quantum numbers $I^{\alpha,\gamma}_j$ is 
given by the $N_{\alpha,\gamma}$. The numbers of 
available $I^{\alpha,\gamma}_j$ is given by the set
$\{(I_j^{\alpha,\gamma})^{min},(I_j^{\alpha,\gamma})^{min}+1,$
$\ldots,(I_j^{\alpha,\gamma})^{max}\}$ and
$(I_j^{\alpha,\gamma})^{min}=-(I_j^{\alpha,\gamma})^{max}$, except for 
$c$, where umklapp processes can break this symmetry. The column
$\phi^{stat}_s$
indicates the possible statistical fluxes.}
\label{uinftydet}
\end{table}
%%%%%%%%%%%%%%%%%%%%%%%%%%%end table%%%%%%%%%%%%%%%%%

%%%%%%%%%%%%%%%%%%%%%%%%%%%end table%%%%%%%%%%%%%%%%%
%%%%%%%%%%%%%%%%%%%%%%%%%%%%%%%%TABLE 4%%%%%%%%%%%%%%%%%
\begin{table}
\begin{tabular}{ccccccccccc}
states & $(I_j^{c})^{min}$&$N_{c}$ 
       & $(I_j^{c,\gamma})^{min}$&$N_{c,\gamma}$ 
	 & $(I_j^{s,0})^{min}$&$N_{s,0}$ 
	 & $(I_j^{s,1})^{min}$&$N_{s,1}$
	 & $(I_j^{s,2})^{min}$&$N_{s,2}$
	 \\
\tableline
Ex.$_0$ & -17/2 & 6&--&--&-1&3& --&--&--&--     \\
Ex.$_{s,1}$ & -8(-9)& 6&--&--& -1&1&0&1&--&--\\
Ex.$_{s,2}$&  -17/2&6&--&--&--&--&--&--&0&1\\
Ex.$_{(c,1;s,1)}$&-8(-9)&4&-6&1&--&--&0&1&--&--\\
Ex.$_{c,1}$&-17/2&4&-6&1&-1/2&2&--&--&--&--\\
Ex.$_{c,1}$&-17/2&2&-13/2&2&0&1&--&--&--&--\\
Ex.$_{c,1}$&--&--&-7&3&--&--&--&--&--&--\\
Ex.$_{c,2}$&-8(-9)&2&-6&1&0&1&--&--&--&--\\
Ex.$_{c,3}$&--&--&-6&1&--&--&--&--&--&--\\
\end{tabular}
\caption{The notation is as follows:  Ex.$_0$, Ex.$_{s,1}$, 
Ex.$_{s,2}$, and Ex.$_{(c,1;s,1)}$ stand for the states 
with no spin strings, the eigenstates with 
one $\gamma=1$ spin-string,  the eigenstates with 
one $\gamma=2$ spin-string, and
the eigenstates with one $\gamma=1$ charge-
and one $\gamma=1$ spin-strings, respectively. Ex.$_{c,\gamma}$ stands for
states with only charge-
strings.
The system size is $L=18$ and the number of electrons
is $N_{\uparrow}=N_{\downarrow}=3$. The number of occupied
quantum numbers $I^{\alpha,\gamma}_j$ is 
given by the $N_{\alpha,\gamma}$. The numbers of 
available $I^{\alpha,\gamma}_j$ is given by the set
$\{(I_j^{\alpha,\gamma})^{min},(I_j^{\alpha,\gamma})^{min}+1,$
$\ldots,(I_j^{\alpha,\gamma})^{max}\}$ and
$(I_j^{\alpha,\gamma})^{min}=-(I_j^{\alpha,\gamma})^{max}$, except for 
$c$, where unklapp processes can break this symmetry.}
\label{tabUt1}
\end{table}
%%%%%%%%%%%%%%%%%%%%%%%%%%%end table%%%%%%%%%%%%%%%%%


\begin{references}
\bibitem{comment} Note that Eq.~(\ref{eq:eigen}) is invariant under
                  the transformation $t\rightarrow -t$,
                  $N_h+N_d \rightarrow L-N_h-N_d$ and $\{k\} \rightarrow
                  C(\{k\})$ (complementar set of wavenumbers).
\end{references}
\end{document}